\documentclass[
    prl, twocolumn, superscriptaddress, nofootinbib, amsmath, amssymb,
    aps, floatfix
]{revtex4-1}

\usepackage{graphicx,xcolor,setspace}
\usepackage{siunitx}
\usepackage{color,epsfig}
\usepackage{ifpdf}
\usepackage{amsmath}
\usepackage{bm}
\usepackage[english]{babel}
\usepackage{amsfonts}%
\usepackage{amssymb}
\usepackage{braket}
\usepackage{enumerate}
\usepackage{multirow}

\usepackage{ulem}
\usepackage{array}
\usepackage{xcolor}

\definecolor{nicered}{rgb}{0.7,0.1,0.1}
\definecolor{nicegreen}{rgb}{0.1,0.5,0.1}

\usepackage{dcolumn}
\usepackage{bm}
\usepackage{slashed}

\begin{document}

\title{Electroweak Precision Fit and New Physics in light of $W$ Boson Mass}


\author{Chih-Ting Lu}
\email{timluyu@kias.re.kr}
\affiliation{Department of Physics and Institute of Theoretical Physics, Nanjing Normal University, Nanjing, 210023, China}

\author{Lei Wu}
\email{leiwu@njnu.edu.cn}
\affiliation{Department of Physics and Institute of Theoretical Physics, Nanjing Normal University, Nanjing, 210023, China}

\author{Yongcheng Wu}
\email{ywu@okstate.edu}
\affiliation{Department of Physics and Institute of Theoretical Physics, Nanjing Normal University, Nanjing, 210023, China}

\author{Bin Zhu}
\email{zhubin@mail.nankai.edu.cn}
\affiliation{Department of Physics, Yantai University, Yantai 264005, China}

\begin{abstract}

The $W$ boson mass is one of the most important electroweak precision observables for testing the Standard Model or its extensions. The very recent measured $W$ boson mass at CDF shows about $7\sigma$ deviations from the SM prediction, which may challenge the internal consistency of the SM. By performing the global electroweak fit with the new $W$ boson,  we present the new values of the oblique parameters: $S = 0.06 \pm 0.10$, $T= 0.11 \pm 0.12$, $U=0.13 \pm 0.09$, or $S=0.14 \pm 0.08$, $T= 0.26 \pm 0.06$ with $U =0$ and the corresponding correlation matrices, which strongly indicates the need for the non-degenerate multiplets beyond the SM.   As a proof-of-concept, we show that the new results can be accommodated in the two-Higgs doublet model, where the charged Higgs boson has to be either heavier or lighter than both two heavy neutral Higgs bosons. Therefore, searching for these non-SM Higgs bosons will provide a complementary way to test the new physics for the $W$ boson mass anomaly.
\end{abstract}

\maketitle


\newpage
\section{Introduction}
\label{sec1}

The electroweak precision observables can assess the validity of the SM~\cite{Hollik:1988ii} but also provide a sensitive probe of uncovering new physics beyond the SM~\cite{Langacker:1991an,Langacker:1991zr}. Historically, the precision experiments at LEP have established or supported the framework of renormalizable gauge field theories. Then, the global fit of the SM to the electroweak precision data indirectly predicted the top quark and the Higgs boson masses before their respective discoveries at the Tevareon and LHC~\cite{ALEPH:2005ab}. Now, in the era of post Higgs boson, the new particles have not been discovered at the LHC, but there exist some intriguing tensions in the precision measurements, such as the muon anomalous magnetic moment and the CKM unitarity triangle. Therefore, as history happened, we may again observe the hints of new physics first in EW precision measurements~\cite{Baak:2012kk,Baak:2014ora,Erler:2019hds}.

In the framework of the global EW fit, the predicted $W$ boson mass is given by~\cite{ParticleDataGroup:2016lqr},
\begin{equation}
m^{\rm SM}_W=80.357 \pm 0.006~{\rm GeV},
\label{eq:mw_fit}
\end{equation}
whose precision is better than that of previous direct measurements at LEP, Tevatron, and LHC. Thus, the sensitivity of EW fit on new physics is limited by the experimental precision of $m_W$~\cite{Baak:2014ora}. Therefore, improving the precision of the direct measurement of $m_W$ is essential for the test of the SM. Very recently, with an unprecedented precision (accuracy of $\sim 1 \times 10^{-4}$), the CDF collaboration at Fermilab has reported the world's best direct measurement of the $W$ boson mass~\cite{CDF:2022hxs},
\begin{equation}
m^{\rm CDF}_W= 80.4335 \pm 0.0094 ~{\rm GeV},
\label{eq:mw_cdf}
\end{equation}
which is on the same level of precision with the EW fits and shows about $7\sigma$ deviation from Eq.~\ref{eq:mw_fit}.

The $W$ boson mass in the SM is related with the $Z$-boson mass, $m_Z$, the fine structure constant, $\alpha$, and the Fermi constant, $G_\mu$ by
\begin{equation}
    m^2_W(1-\frac{m^2_W}{m^2_Z})=\frac{\pi\alpha}{\sqrt{2}G_\mu}(1+\Delta r).
    \label{eq:mw_th}
\end{equation}
Here $\Delta r$ includes the quantum corrections to $m_W$, which depends on the top quark mass, $m_t$, quadratically and the Higgs mass, $m_h$, logarithmically. Since the relation in Eq.~\ref{eq:mw_th} is of central importance to the precision tests of the electroweak sector of the SM, the theoretical prediction of $m_W$ in the SM has been calculated at the one-loop~\cite{Sirlin:1980nh,Marciano:1980pb} and two-loop level~\cite{Djouadi:1987gn,Djouadi:1987di,Kniehl:1989yc,Halzen:1990je,Kniehl:1991gu,Freitas:2000gg,Freitas:2002ja,Awramik:2002wn,Awramik:2003ee,Onishchenko:2002ve,Awramik:2002vu}, as well as leading three- and four-loop corrections~\cite{Avdeev:1994db,Chetyrkin:1995ix,Chetyrkin:1995js,Chetyrkin:1996cf,Faisst:2003px,vanderBij:2000cg,Boughezal:2004ef,Boughezal:2006xk,Chetyrkin:2006bj,Schroder:2005db}. The $m_W$ containing all known higher-order corrections in the on-shell scheme has been given in Refs.~\cite{Awramik:2006uz,Degrassi:2014sxa}. The unknown higher order contributions have been estimated to be about 4-6 MeV~\cite{Awramik:2003rn,Degrassi:2014sxa}. Then, including these theoretical corrections, there is still about $5.1\sigma$ discrepancy from  Eq.~\ref{eq:mw_fit}. This may be an under-estimation of the high order corrections or systematic error, or a hint of the new physics beyond the SM.


In this work, we first calculate the theoretical values of EWPOs by performing the global EW fit with the newly measured $W$ boson mass and show the new tensions between the EW fit results and the direct measurements. Then we derive the model-independent constraints on the new physics, such as the oblique parameters $S$, $T$ and $U$, and discuss the implications for new physics models. Finally, we draw our conclusions.

\section{Electroweak Fits}
\label{sec2}

\begin{table*}[ht]
\caption{The input parameters and the best points in the global EW fit. The Fermi constant $G_F = 1.1663787(6)\times 10^{-5}$ [GeV$^{-2}$]~\cite{ParticleDataGroup:2020ssz} is fixed in our calculation. Correlations among $(m_Z,\Gamma_Z,\sigma_h^0,R_\ell^0,A_{\rm FB}^{0,\ell})$ and among $(A_{\rm FB}^{0,c},A_{\rm FB}^{0,b},A_c,A_b,R_c^0,R_b^0)$ are also taken into account~\cite{ALEPH:2005ab}. The value of ``Pull'' is defined as $(O_{\rm fit} - O_{\rm measure})/\sigma_{\rm measure}$, where $\sigma_{\rm measure}$ is the error of each input observable.}
\label{tab:fit}
    {\renewcommand{\arraystretch}{1.2}
    \scalebox{0.8}{
    \begin{tabular}{l|l|lr|lr|lr|lr|r}
    \hline
    \multirow{3}{*}{Parameter}                                       & \multirow{3}{*}{Input Value}            &   \multicolumn{4}{c|}{PDG 2021}  & \multicolumn{4}{c|}{CDF 2022} & \multirow{3}{*}{Refs} \\\cline{3-10}
    & & \multicolumn{2}{c|}{$\chi^2_{\rm min}(\rm dof) = 18.73(16)$} & & & \multicolumn{2}{c|}{$\chi^2_{\rm min}(\rm dof)= 64.45(16)$} & & & \\
    & & Fit Result & Pull & Fit w/o Input & Pull & Fit Result & Pull & Fit w/o Input & Pull &\\
    \hline\hline
    \multirow{2}{*}{$m_W$ [GeV]}                                     &        $80.379(12)$               &  $80.361(6)$ & $-1.47$~ & $80.357(6)$ & $-1.86$~  & --  & -- &   -- & --~  &  \cite{ParticleDataGroup:2020ssz}         \\
    &  $80.4335(94)$ & -- & -- & -- & --~ & $80.381(5)$ & $-5.80$~ & $80.357(6)$ & $-8.53$~ & \\
    \hline\hline
    $\Delta \alpha_{\rm had}^{(5)}$\footnote{Scaled with $\alpha_s(m_Z)$.}                 & $0.02761(11)$   &     $0.02756(11)$ & $-0.44$~  & $0.02716(38)$ & $-4.06$~   &  $0.02746(10)$ & $-1.37$ & $0.02603(36)$  & $-14.37$~ &   \cite{Crivellin:2020zul,Davier:2019can,Keshavarzi:2019abf} \\
    $m_h$  [GeV]                                    & $125.25(17)$  &  $125.25(17)$ & $-0.02$~ & $92^{(21)}_{(18)}$ & $-193.26$~  &  $125.24(17)$ & $-0.06$ & $42^{(10)}_{(8)}$  & $-489.71$~ &  \cite{ParticleDataGroup:2020ssz}         \\
    $m_t$ [GeV]\footnote{$0.5$ GeV theoretical uncertainty is included.} & $172.76(58)$  & $173.02(56)$ & $0.45$~ & $176.2(20)$ & $5.83$~ &  $174.04(55)$ & $2.19$~ & $184.2(16)$ & $19.55$~   &  \cite{ParticleDataGroup:2020ssz}         \\
    $\alpha_s(m_Z)$                                 & $0.1179(9)$ & $0.1180(9)$ & $0.14$~ & $0.1193(9)$ & $1.53$~ &  $0.1177(9)$ & $-0.26$~ & $0.1152(29)$ & $-0.22$~  &  \cite{ParticleDataGroup:2020ssz}         \\
    $\Gamma_W$ [GeV]                                & $2.085(42)$ & $2.0905(5)$ & $0.13$~ & $2.0905(5)$ & $0.13$~ &  $2.0919(5)$ & $0.16$~ & $2.919(5)$  & $0.16$~  &   \cite{ParticleDataGroup:2020ssz}        \\
    $\Gamma_Z$ [GeV]                                & $2.4952(23)$ & $2.4942(6)$ & $-0.45$~ & $2.4940(7)$ & $-0.51$~ &  $2.4946(6)$ & $-0.26$~ & $2.4945(7)$  & $-0.31$~   &  \cite{ALEPH:2005ab}         \\
    $m_Z$ [GeV]                                     & $91.1875(21)$ & $91.1882(20)$ & $0.34$~ & $91.2037(90)$ & $7.72$~ &  $91.1909(20)$ & $1.63$~ & $91.2393(77)$  & $24.66$~  &   \cite{ALEPH:2005ab}        \\
    $A_{\rm FB}^{0,b}$                              & $0.0992(16)$ & $0.1031(3)$ & $2.44$~ & $0.1033(3)$ & $2.54$~ &  $0.1036(3)$ & $2.72$~ & $0.1037(3)$   &  $2.83$~ &   \cite{ALEPH:2005ab}        \\
    $A_{\rm FB}^{0,c}$                              & $0.0707(35)$ & $0.0737(3)$ & $0.85$~ & $0.0737(3)$ & $0.85$~ &  $0.0740(3)$ & $0.95$~ & $0.07404(25)$ & $0.95$~  &   \cite{ALEPH:2005ab}        \\
    $A_{\rm FB}^{0,\ell}$                           & $0.0171(10)$  & $0.01623(10)$ & $-0.87$~ & $0.01622(10)$ & $-0.88$~ & $0.01637(10)$ & $-0.73$~ & $0.01636(10)$ & $-0.74$~  &   \cite{ALEPH:2005ab}        \\
    $A_b$                                           & $0.923(20)$ & $0.93462(4)$ & $0.58$~ & $0.93462(4)$ & $0.58$~ &  $0.93464(4)$ & $0.58$~ & $0.93464(4)$  & $0.58$~    &  \cite{ALEPH:2005ab}         \\
    $A_c$                                           & $0.670(27)$  & $0.6679(2)$ & $-0.08$~ & $0.6679(2)$ & $-0.08$~ &  $0.6682(2)$ & $-0.07$ & $0.6682(2)$ & $-0.07$~    &  \cite{ALEPH:2005ab}         \\
    $A_\ell(\rm SLD)$                               & $0.1513(21)$ & $0.1471(5)$ & $-2.00$~ & $0.1469(5)$ & $-2.10$~ &  $0.1478(5)$ & $-1.70$ & $0.1476(5)$ & $-1.78$~   &  \cite{ALEPH:2005ab}         \\
    $A_\ell(\rm LEP)$                               & $0.1465(33)$ & $0.1471(5)$ & $0.18$~ & $0.1469(5)$ & $0.12$~ &  $0.1478(5)$ & $0.37$~ & $0.1476(5)$ & $0.32$~   &  \cite{ALEPH:2005ab}         \\
    $R_b^0$                                         & $0.21629(66)$ & $0.21583(10)$ & $-0.69$~ & $0.21582(10)$ & $-0.71$~ & $0.21580(10)$ & $-0.74$ & $0.21579(10)$ & $-0.76$~  &  \cite{ALEPH:2005ab}         \\
    $R_c^0$                                         & $0.1721(30)$ & $ 0.17222(6)$ & $0.04$~ & $0.17222(6)$ & $0.04$~ &  $0.17223(6)$ & $0.04$ & $0.17223(6)$ & $0.04$~    &  \cite{ALEPH:2005ab}         \\
    $R_\ell^0$                                      & $20.767(25)$ & $ 20.735(8)$ & $-1.28$~ & $20.732(8)$ & $-1.40$~ &   $20.733(8)$ & $-1.35$ & $20.730(8)$ & $-1.48$~    &   \cite{ALEPH:2005ab}        \\
    $\sigma_h^0$ [nb]                               & $41.540(37)$ & $41.491(8)$ & $-1.34$~ & $41.489(8)$ & $-1.39$~ & $41.490(8)$ & $-1.35$ & $41.488(8)$  & $-1.39$~    &   \cite{ALEPH:2005ab}        \\
    $\sin^2\theta_{\rm eff}^{\ell}(Q_{FB})$         & $0.2324(12)$ & $0.23151(6)$ & $-0.74$~  & $0.23151(6)$ & $-0.74$~ & $0.23143(6)$ & $-0.81$~  & $0.23143(6)$ & $-0.81$~  &  \cite{ALEPH:2005ab}         \\
    $\sin^2\theta_{\rm eff}^{\ell}({\rm Teva})$ & $0.23148(33)$ & $0.23151(6)$ & $0.10$~ & $0.23151(6)$  & $0.10$~ & $0.23143(6)$ & $-0.15$ & $0.23143(6)$ & $-0.15$~ &  \cite{CDF:2018cnj}        \\
    $\overline{m}_c$ [GeV]                          & $1.27(2)$ & $1.27(2)$ & $0.00$~ & -- & --~ & $1.27(2)$ & $0.00$~ & -- & --~ &  \cite{ParticleDataGroup:2020ssz}         \\
    $\overline{m}_b$ [GeV]                          & $4.18^{(3)}_{(2)}$ & $4.18^{(3)}_{(2)}$ & $0.00$~ & -- & --~ & $4.18^{(3)}_{(2)}$ & $0.00$ & -- & --~ &  \cite{ParticleDataGroup:2020ssz}          \\
    \hline\hline
    \end{tabular}}
    }
\end{table*}

The global EW fit is a powerful tool to explore the correlations among observables in the SM and predict the direction of new physics~\cite{Flacher:2008zq,Baak:2011ze,Baak:2012kk,Baak:2014ora,Haller:2018nnx,Crivellin:2020zul,DeBlas:2019ehy}. Since the EW parameters in the SM are closely related to each other, we can expect some observables in the global EW fits may suffer from new tensions once the $m_W$ is changed. We use {\tt Gfitter}~\cite{Flacher:2008zq,Baak:2011ze,Baak:2012kk,Baak:2014ora,Haller:2018nnx} with data from Refs.~\cite{ParticleDataGroup:2020ssz,ALEPH:2005ab,CDF:2018cnj,Crivellin:2020zul,Davier:2019can,Keshavarzi:2019abf} and two benchmark $m_W$ values : (1) $80.379\pm 0.012$ GeV [PDG (2021)], (2) $80.4335\pm 0.0094$ GeV [CDF (2022)] to investigate the variations of these observables. The numerical results are presented in the Tab.~\ref{tab:fit} and Fig.~\ref{fig:pull}.

It can be found that the PDG (2021) has $\chi^2_{min}(dof)=18.73(16)$, which is generally in good agreement with the SM predictions. However, the new CDF (2022) has $\chi^2_{min}(dof)=64.45(16)$, which means the sizable discrepancies between the best points from EW fits and input parameters, especially for $m_W$, $m_t$, $m_Z$ and the hadronic contribution to the shift in the fine structure constant $\Delta\alpha^{(5)}_{had}$ compared with the PDG (2021) ones.

To be specific, the relation between $m_W$ and $m_Z$ in the $\overline{MS}$ scheme can be written as $m_W = m_Z\rho^{1/2}c_W$ where $\rho\sim 1+\frac{3G_F m^2_t}{8\sqrt{2}\pi^2}$ and $c_W\equiv \sqrt{1-\sin^2\theta_W (m_Z)}$ is the cosine of the Weinberg angle. As shown in Tab.~\ref{tab:fit}, the global EW fits of $\sin^2\theta_{\rm eff}^{\ell}$ is consistent with the measured value, so we can only increase the $m_Z$ and $m_t$ to enhance the $m_W$. However, even the differences between the best points of $m_Z$ and $m_t$ are already about $2\sigma$ from the input parameters, the best points still cannot reach to the new CDF measured $m_W$. This explains the large and negative Pull in the second row of Tab.~\ref{tab:fit}. We also note that there was a $2.8\sigma$ discrepancy between the two most precise top quark mass measurements, $174.98\pm 0.76$ GeV (D$\emptyset$)~\cite{D0:2014yem} and $172.25\pm 0.63$ GeV (CMS)~\cite{CMS:2018quc}. While the global EW fit with the new CDF (2022) predicts the heavier top quark mass. From $m^2_W$ in Eq.~\ref{eq:mw_th}, we can find the $m^2_W$ is anti-correlated with the fine structure constraint. Hence, the enhanced $m_W$ value reported in CDF (2022) is related to the smaller $\Delta\alpha^{(5)}_{had}$ in the EW fits.
In addition, the decrease of $\Delta\alpha^{(5)}_{had}$ can be translated to the smaller hadronic vacuum polarization contributions to the muon anomalous magnetic moment, $a^{HVP}_{\mu}$~\cite{Keshavarzi:2020bfy,deRafael:2020uif}. Therefore, the difference between $a_{\mu}(\rm Exp)$ and $a_{\mu}(\rm SM)$ can be enlarged if the $a^{HVP}_{\mu}$ is extracted from $\Delta\alpha^{(5)}_{had}$ of the global EW fits. Besides, the old tension for $A^{0,b}_{\rm FB}$ ($A_l$) in PDG (2021) is increased (decreased) in the EW fits with new CDF measured $m_W$. Hence, the measurement of $A^{0,b}_{\rm FB}$ needs special treatment in the future. On the other hand, we also show the predictions of each observable by removing its input value once a time in the EW fits in the fourth and sixth columns in the Tab.~\ref{tab:fit}.
As expected, $m_t$, $m_Z$ and $\Delta\alpha^{(5)}_{had}$ are sensitive to the change of $m_W$. Moreover, if the Higgs mass $m_h$ measured by LHC is removed, its best point from the global EW fits is dramatically reduced. It is because the $W$ boson mass can be written as~\cite{Awramik:2003rn}
\begin{equation}
m_W = m^0_W - C_1 \ln r_h + C_2 (r^2_t - 1) - C_3 \ln r_h (r^2_t - 1) + ...
\label{eq:mW_para}
\end{equation}
where $m^0_W$ is the leading order value of $W$ boson mass, and $r_h\equiv m_h/(100~\text{GeV})$, $r_t\equiv m_t/(173.4~\text{GeV})$. $C_1$, $C_2$ and $C_3$ are positive coefficients. Once the $m_W$ is increased, the prediction of $m_h$ without its input values in the EW fits will be decreased to compensate the difference between $m_W$ and $m^0_W$. Note that without the LHC input for $m_h$, the CDF (2022) measurement of $m_W$ together with other EWPOs indicates an extremely light Higgs $m_h\approx 42^{+10}_{-8}$ GeV, which is considerably inconsistent with current measurement. If the CDF (2022) measurement is confirmed by other experiments, it strongly indicates that there is unknown correction to $m_W$ from $m_h$ in SM or there is new physics in the scalar sector. When both $m_t$ and $m_h$ are not used in the EW fits, the antagonistic effect between $m_t$ and $m_h$ makes the allowed regions oblique.
In order to fit the $m_W$ value and minimize the total $\chi^2$, heavier $m_t$ and $m_h$ are preferred as the best point. The above numerical results are visualized in the Fig.~\ref{fig:1D_ewfit} and ~\ref{fig:2D_ewfit} in supplementary materials. Besides, we also use both values of $m_W$ in PDG (2021) and CDF (2022) as the input parameters in the global EW fits, but find that the best points are just slightly different from that only using the CDF (2022).




\section{New Bounds and New Physics}
\label{sec3}

\begin{table}[ht]
    \caption{The values of $S$, $T$ and $U$ and the correlation matrix allowed by the EW fit with the $W$ boson mass from CDF (2022) and PDG (2021), respectively. $m_h=125$ GeV and $m_t=172.5$ GeV are used as the SM reference point.}
    \label{tab:stu}
    \resizebox{0.5\textwidth}{!}{
    \begin{tabular}{l|rrrr|rrrr}
    \hline
    \multirow{3}{*}{$13\,\rm dof$} & \multicolumn{4}{c|}{PDG 2021 }                                                                                                      & \multicolumn{4}{c}{CDF 2022}                                                                                                                    \\ \cline{2-9}
                      & \multicolumn{1}{c|}{Result}                  & \multicolumn{3}{c|}{Correlation}                                                                  & \multicolumn{1}{c|}{{Result}}                & \multicolumn{3}{c}{Correlation}                                                                  \\
                      & \multicolumn{1}{c|}{$\chi^2_{\rm min} = 15.42$}  & \multicolumn{1}{c}{$S$} & \multicolumn{1}{c}{$T$} & \multicolumn{1}{c|}{$U$} & \multicolumn{1}{c|}{$\chi^2_{\rm min} = 15.44$}                    & \multicolumn{1}{c}{$S$} & \multicolumn{1}{c}{$T$} & \multicolumn{1}{c}{$U$} \\ \hline
    $S$        & \multicolumn{1}{r|}{$0.06\pm0.10$}           & $1.00$                         & $0.90$                         & $-0.57$                         & \multicolumn{1}{r|}{$0.06\pm0.10$}           & $1.00$                         & $0.90$                         & $-0.59$                        \\
    $T$        & \multicolumn{1}{r|}{$0.11\pm0.12$}           &                                & $1.00$                         & $-0.82$                         & \multicolumn{1}{r|}{$0.11\pm0.12$}           &                                & $1.00$                         & $-0.85$                        \\
    $U$        & \multicolumn{1}{r|}{$-0.02\pm0.09$}          &                                &                                & $1.00$                          & \multicolumn{1}{r|}{$0.14\pm0.09$}           &                                &                                & $1.00$                         \\ \hline
    \end{tabular}
    }
\end{table}

\begin{table}[ht]
    \caption{Same as Tab.~\ref{tab:stu}, but for $S$ and $T$ with $\Delta U=0$.}
    \label{tab:st}
    \resizebox{0.5\textwidth}{!}{
    \begin{tabular}{l|ccr|ccr}
    \hline
    \multirow{2}{*}{$U = 0$} & \multicolumn{3}{c|}{PDG 2021}                                                                 & \multicolumn{3}{c}{CDF 2022}                                                                               \\ \cline{2-7}
                      & \multicolumn{1}{c|}{{Result}}                & \multicolumn{2}{c|}{Correlation}                             & \multicolumn{1}{c|}{{Result}}                & \multicolumn{2}{c}{Correlation}                             \\
    $14\,\rm dof$     & \multicolumn{1}{c|}{$\chi^2_{\rm min}=15.48$}& $S$                 & \multicolumn{1}{c|}{$T$} & \multicolumn{1}{c|}{$\chi^2_{\rm min}=17.82$}& $S$                 & \multicolumn{1}{c}{$T$} \\ \hline
    $S$        & \multicolumn{1}{r|}{$0.05\pm0.08$}           & \multicolumn{1}{r}{$1.00$} & $0.92$                          & \multicolumn{1}{r|}{$0.15\pm0.08$}           & \multicolumn{1}{r}{$1.00$} & $0.93$                         \\
    $T$        & \multicolumn{1}{r|}{$0.09\pm0.07$}           & \multicolumn{1}{r}{  }     & $1.00$                          & \multicolumn{1}{r|}{$0.27\pm0.06$}           & \multicolumn{1}{r}{  }     & $1.00$                         \\ \hline
    \end{tabular}
    }
\end{table}

\begin{figure}[ht]
    \includegraphics[width=0.45\textwidth]{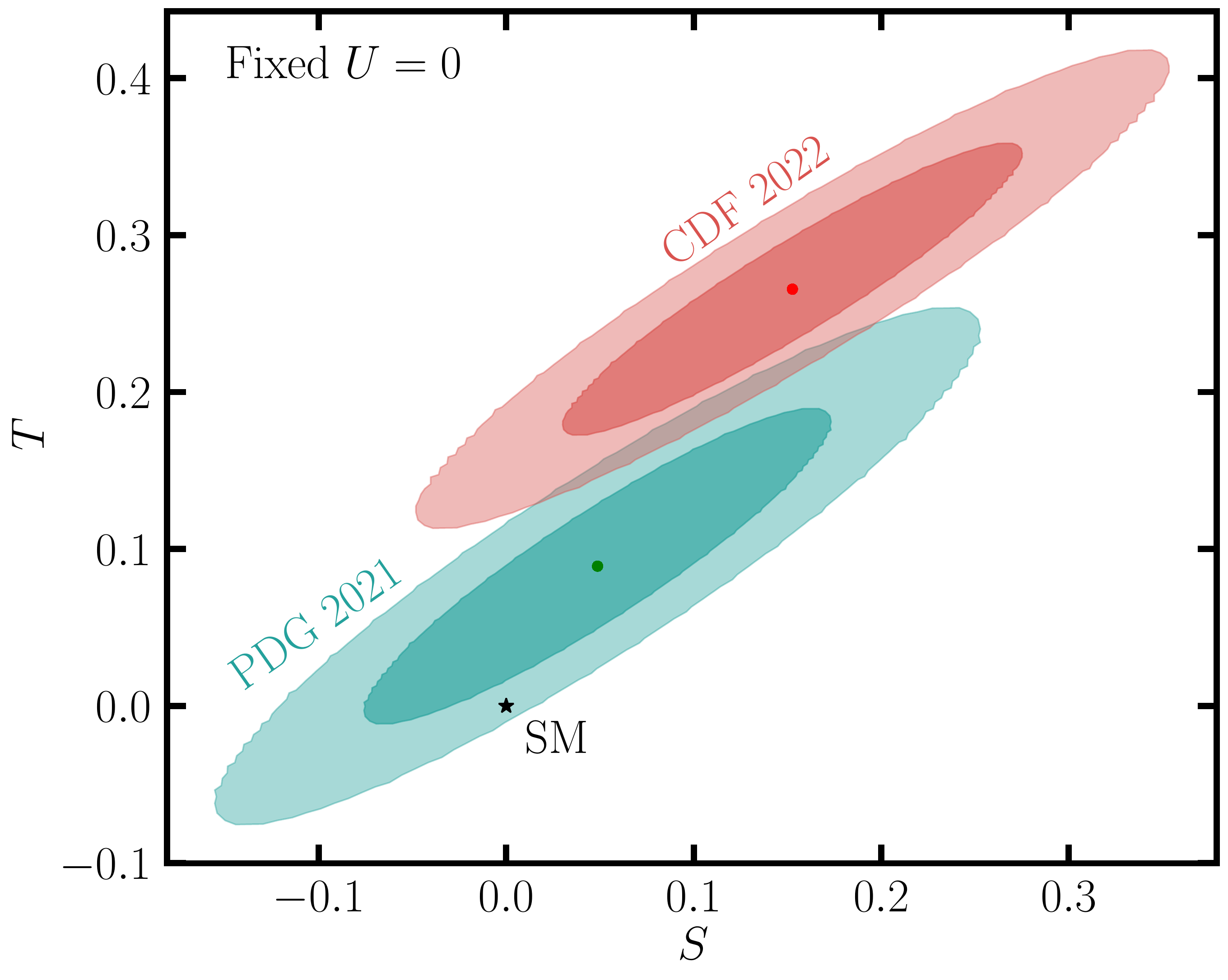}
    \caption{The 1- and 2-$\sigma$ allowed regions in $S$-$T$ plane from the electroweak fits using the PDG 2021 data set with the old value of $m_W$ (green region) and the new CDF value of $m_W$ (red region).}
    \label{fig:s_t_ewfit}
\end{figure}

The EWPOs generically impose stringent constraints on any theory of electroweak symmetry breaking.  Most of the new physics effects on precision measurements can be described by the oblique parameters $S$, $T$, and $U$\footnote{Besides, there are other equivalent constraints from the EWPOs, such as ($M_W$, $\rho$, $\sin^2\theta_{eff}$) and ($\epsilon_1$, $\epsilon_2$, $\epsilon_3$) parameters~\cite{Altarelli:1990zd,Altarelli:1991fk,Barbieri:1991qp,Altarelli:1993bh,Burgess:1993vc}.}~\cite{Peskin:1991sw}. In~Tab.~\ref{tab:stu}, we give the allowed values of $S$, $T$ and $U$ and the correlation matrix by using the EW fit with the $W$ mass from CDF (2022) and PDG (2021), respectively. The main differences are that the central value of $U$ parameter predicted by CDF (2022) is much larger than that predicted by PDG (2021), and the correlations between $U$ and $S$ or $T$ are mildly strengthened as well. If making $U > S,T$, one may need to introduce some new large multiplet with sufficient low masses of the components beyond the SM~\cite{Lavoura:1993nq}. We note that the $\chi^2_{min}$ in~Tab.~\ref{tab:fit} can be reduced to $15.44$ from $64.45$ if including $S$, $T$ and $U$ in the fit, which demonstrate that the oblique parameters can describe the main effects caused by newly measured $W$ boson mass. On the other hand, since the values of $U$ parameter are found to be very small in many new physics models, we also present the results for $S$ and $T$ with $\Delta U=0$ in Tab.~\ref{tab:st} and the Fig.~\ref{fig:s_t_ewfit}. Without the extra freedom of the $U$ parameter, one can only increase both $S$ and $T$ parameters to fit $m_W$. It can be seen that the SM value is within the $2\sigma$ allowed region by the PDG (2021), however, which is far away from that given by CDF (2022). The overlap between PDG (2021) and CDF (2022) results only appears in the $2\sigma$ allowed region in the $S$-$T$ plane. Hence, if interpreted in the new physics, the newly measured $m_W$ by CDF may favor introducing the additional multiplets beyond the SM. With this observation, we take the two Higgs doublet model (2HDM) as an example to show the parameter space allowed by new oblique parameters. Note that there are other experimental constraints, such as flavor observables, Higgs data and LHC bounds. A comprehensive global fit of 2HDM will be presented in the later work~\cite{2hdmew}.

In the 2HDM~\cite{Branco:2011iw}, there are five massive spin-zero states in the spectrum $(h, H, A, H^\pm)$ after the electroweak symmetry breaking (EWSB). As an illustration, we consider the alignment limit in which one of the two neutral CP-even Higgs mass eigenstates aligns with the direction of the scalar field vacuum expectation values (vevs). We assume $m_h=125$ GeV and then the alignment limit corresponds to $\cos(\beta-\alpha) \to 0$~\cite{Gunion:2002zf,Craig:2013hca,Carena:2013ooa,Bernon:2015qea,Chen:2018shg,Eberhardt:2020dat}. The new contributions to the oblique parameters in the 2HDM arise from the non-SM Higgs loops. It should be noted that the $U$ parameter is much smaller than $S$ and $T$ parameters in our considered parameter space~(c.f.~Fig.~\ref{fig:2hdm_stu_0}). Thus, we use the allowed values of $S$ and $T$ in Tab.~\ref{tab:st} to perform a fit by defining $\chi^2({\bold O})=({\bold y}-{\bold \mu}({\bold O}))^T {\bold C}^{-1} ({\bold y}-{\bold \mu}({\bold O}))$, where $\bold y$ is the vector of central values and $\bold C$ is the covariance matrix.

\begin{figure}[ht]
\includegraphics[width=9cm,height=7cm]{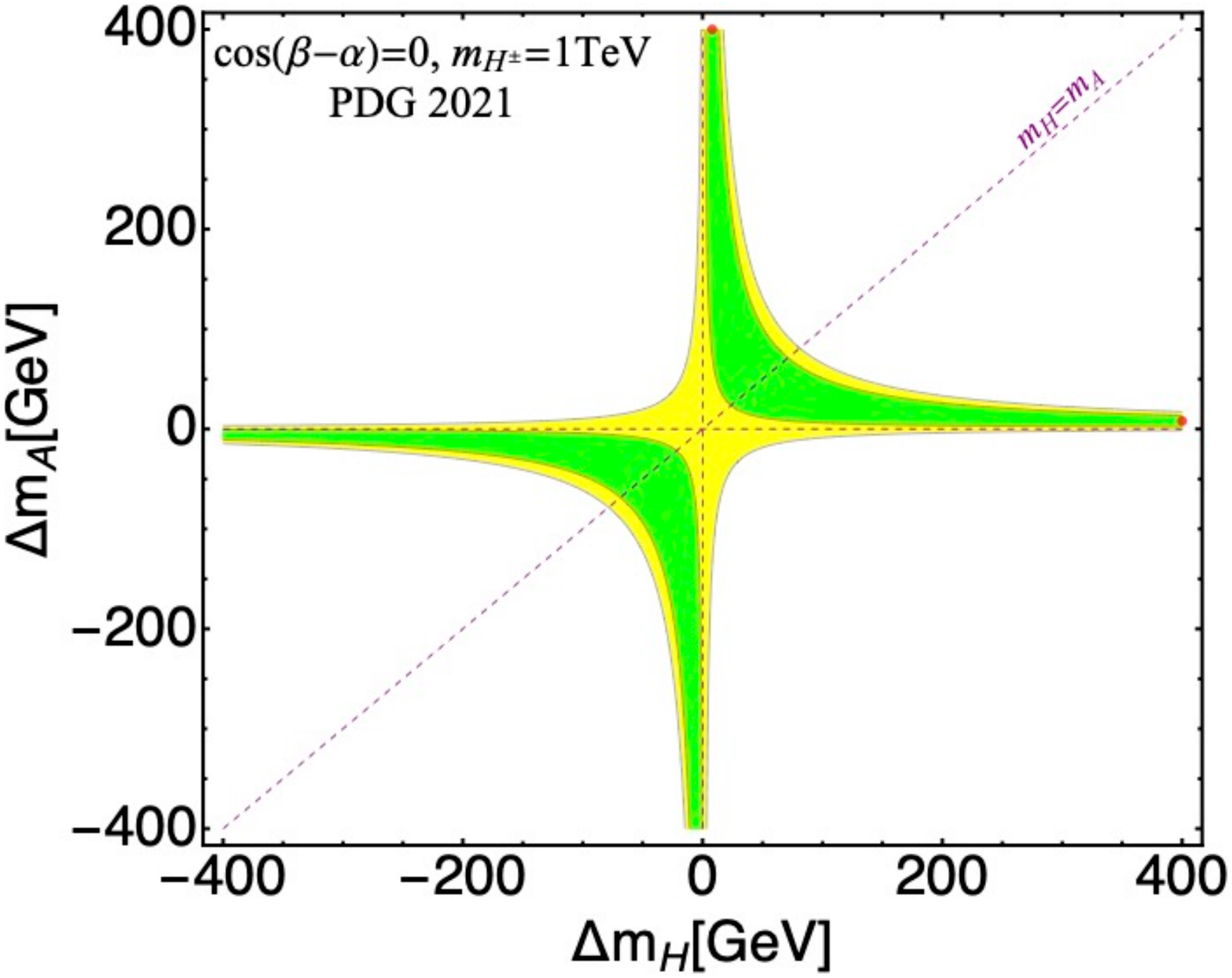}
\includegraphics[width=9cm,height=7cm]{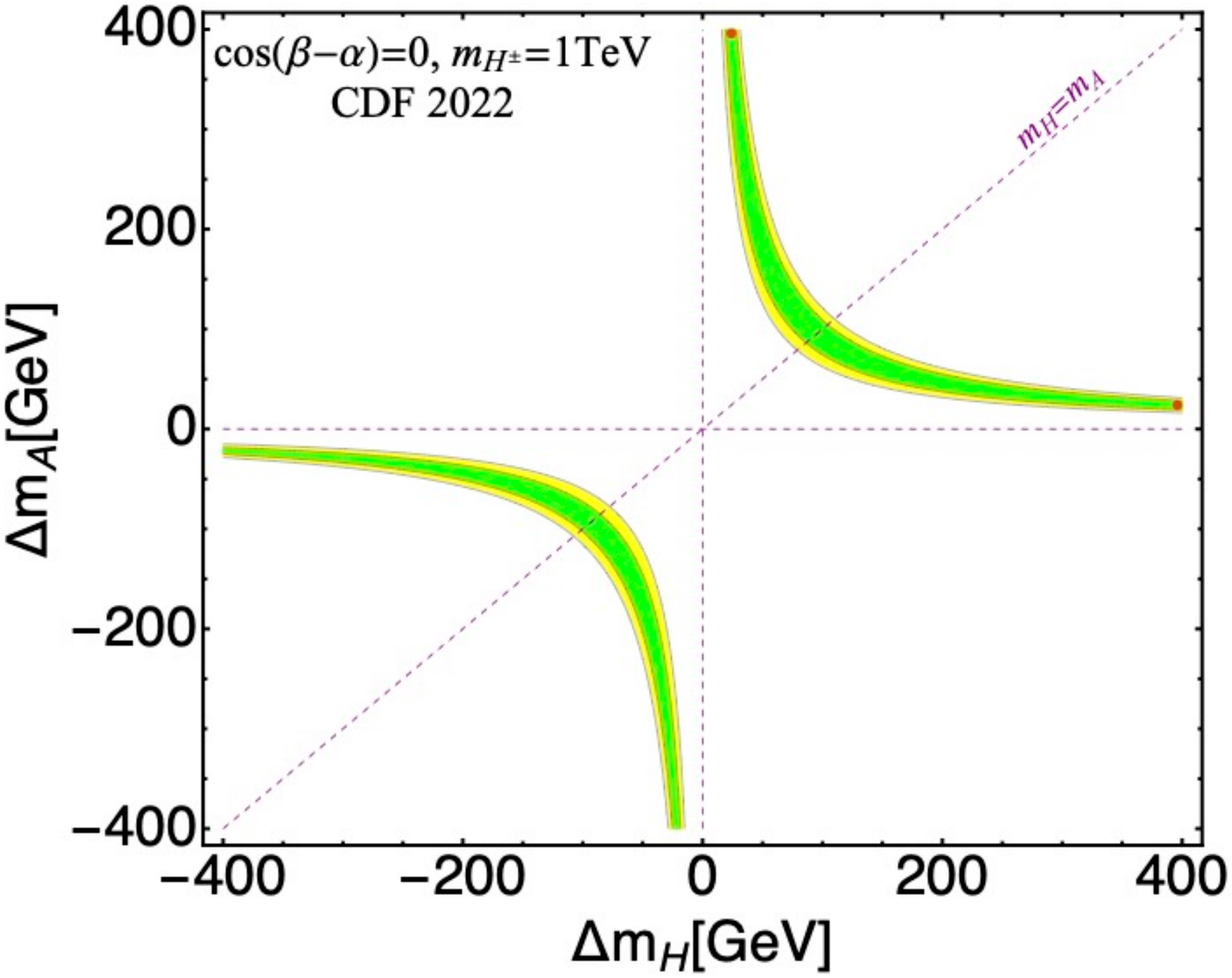}
\caption{The 1- and 2-$\sigma$ regions allowed by the oblique parameter fit for 2HDM in the alignment limit in the plane of $\Delta m_A$ versus $ \Delta m_H$, where $\Delta m_A=m_A - m_{H^\pm}$ and $\Delta m_H=m_H - m_{H^\pm}$. The best points: $(\Delta m_{A},\Delta m_{H})=(24, 396)=(396, 24)$ GeV with $\chi^2_{\min}=3.04$, $S=0.01$ and $T=0.173$ [CDF (2022)]; $(\Delta m_{A},\Delta m_{H})=(8, 400)=(400, 8)$ GeV with $\chi^2_{\min} =0.24$ and $S=0.01$, $T=0.058$ [PDG (2021)], respectively. }
\label{fig:2hdm_bestfit}
\end{figure}

In Fig.~\ref{fig:2hdm_bestfit}, we show 1- and 2-$\sigma$ region allowed by $S$ and $T$ parameters in the plane of $\Delta m_A$-$ \Delta m_H$ using the PDG 2021 data set with the old value of $m_W$ and the new CDF value of $m_W$. We find the results are symmetric about $m_H=m_A$ axis because the $S$ and $T$ parameters are unchanged under the exchange of $m_H$ and $m_A$ (c.f. Eq.~\ref{eq:stu-2hdm}). In the fit, the $T$ parameter is dominant because of its quadratic dependence on the masses of the new particles. However, if the mass splitting between the charged Higgs bosons and neutral Higgs bosons is too large, it will over-enhance the $T$ parameter. Besides, in general, there is a perturbative bound on the mass splittings between these scalars, e.g. $m_{H/A}^2-m_{H^\pm}^2\sim \bar{\lambda}v^2 < 4\pi v^2$, where $v$ is the vev of the Higgs field and $\bar{\lambda}$ is a combination of the quartic couplings in 2HDM~(c.f.~\ref{eqn:mass}). Compared with the result in our PDG (2021) fit, the degenerate case $m_{H^\pm}=m_A=m_H$ in new CDF (2022) fit is strongly disfavored. Besides, when $m_{H^{\pm}}=m_{H}$ or $m_{H^{\pm}}=m_A$, $T$ parameter vanishes because these conditions lead to an exact custodial $SU(2)$ symmetry, namely one of the neutral scalars joins the charged scalars to form an $SU(2)$ triplet. On the other hand, if the charged Higgs mass $m_{H^\pm}$ lies between the masses of the two neutral scalars, $m_A$ and $m_H$, the $T$ parameter is negative. Thus, only $m_{H^\pm} > m_{A},~m_{H}$ or $m_{H^\pm} < m_{A},~m_{H}$ are allowed. However, in the former case, $S$ parameter is inclined to be negative (c.f.~Fig~\ref{fig:2hdm_stu_0}) because of $S \sim {\rm log}(m_{H,A}/m_{H^\pm})$. This makes the best points lie in the latter case. Therefore, the charged Higgs boson mass $m_{H^\pm}$ has to be non-degenerate with two neutral Higgs boson masses $m_A$ and $m_H$. In more general cases, although the Higgs doublets can mix, this will not change our main conclusion (c.f.~Fig.~\ref{fig:2hdm-stu-1}). Depending on the spectrum of the heavy Higgs bosons, one may probe them in some cases at the LHC or future higher energy colliders. For example, if $H^\pm$ is the lightest newHiggs boson, it can be searched for through the process $pp \to tbH^\pm \to t\bar{t}b\bar{b}$~\cite{ATLAS:2021upq,CMS:2020imj}. While if $H$ or $A$ is the lightest new Higgs boson, one can look for them through the process $pp \to t\bar{t} H/A \to t\bar{t}t\bar{t}$~\cite{ATLAS-CONF-2022-008,CMS:2019rvj}.

\section{Conclusion}
\label{sec4}
The very recent measurement result of $m_W$ at CDF deviates from the SM prediction by about $7\sigma$, which leads to deviations in the electroweak fit. Based on our analysis, we find the predicted $Z$ boson mass $m_Z$, the top quark mass $m_t$ and the hadronic contribution to the shift in the fine structure constant $\Delta \alpha^{(5)}_{\rm had}$ also show the mild discrepancies from their respective measurements. Furthermore, we derived the new results of the oblique parameters $S$, $T$ and $U$ from the electroweak fit, 
which strongly imply the SM extensions with new non-degenerate multiplets and their mass sequence in the spectrum. Using 2HDM as an example, we demonstrated that the charged Higgs boson, being non-degenerate with two heavy neutral Higgs bosons is required, and it has to be either heavier or lighter than both of them. The search for these non-SM Higgs bosons can be used to probe the new physics for $W$ boson anomaly at the LHC and future colliders.


\section{Acknowledgements}
\label{sec5}
This work is supported by the National Natural Science Foundation of China (NNSFC) under grant No.~12147228, 11805161. Y.W. would like to thank U.S.~Department of Energy for the financial support, under grant number DE-SC 0016013.

\bibliography{refs}

\begin{thebibliography}{69}
\expandafter\ifx\csname natexlab\endcsname\relax\def\natexlab#1{#1}\fi
\expandafter\ifx\csname bibnamefont\endcsname\relax
  \def\bibnamefont#1{#1}\fi
\expandafter\ifx\csname bibfnamefont\endcsname\relax
  \def\bibfnamefont#1{#1}\fi
\expandafter\ifx\csname citenamefont\endcsname\relax
  \def\citenamefont#1{#1}\fi
\expandafter\ifx\csname url\endcsname\relax
  \def\url#1{\texttt{#1}}\fi
\expandafter\ifx\csname urlprefix\endcsname\relax\def\urlprefix{URL }\fi
\providecommand{\bibinfo}[2]{#2}
\providecommand{\eprint}[2][]{\url{#2}}

\bibitem[{\citenamefont{Hollik}(1990)}]{Hollik:1988ii}
\bibinfo{author}{\bibfnamefont{W.~F.~L.} \bibnamefont{Hollik}},
  \bibinfo{journal}{Fortsch. Phys.} \textbf{\bibinfo{volume}{38}},
  \bibinfo{pages}{165} (\bibinfo{year}{1990}).

\bibitem[{\citenamefont{Langacker and Luo}(1991)}]{Langacker:1991an}
\bibinfo{author}{\bibfnamefont{P.}~\bibnamefont{Langacker}} \bibnamefont{and}
  \bibinfo{author}{\bibfnamefont{M.-x.} \bibnamefont{Luo}},
  \bibinfo{journal}{Phys. Rev. D} \textbf{\bibinfo{volume}{44}},
  \bibinfo{pages}{817} (\bibinfo{year}{1991}).

\bibitem[{\citenamefont{Langacker et~al.}(1992)\citenamefont{Langacker, Luo,
  and Mann}}]{Langacker:1991zr}
\bibinfo{author}{\bibfnamefont{P.}~\bibnamefont{Langacker}},
  \bibinfo{author}{\bibfnamefont{M.-x.} \bibnamefont{Luo}}, \bibnamefont{and}
  \bibinfo{author}{\bibfnamefont{A.~K.} \bibnamefont{Mann}},
  \bibinfo{journal}{Rev. Mod. Phys.} \textbf{\bibinfo{volume}{64}},
  \bibinfo{pages}{87} (\bibinfo{year}{1992}).

\bibitem[{\citenamefont{Schael et~al.}(2006)}]{ALEPH:2005ab}
\bibinfo{author}{\bibfnamefont{S.}~\bibnamefont{Schael}} \bibnamefont{et~al.}
  (\bibinfo{collaboration}{ALEPH, DELPHI, L3, OPAL, SLD, LEP Electroweak
  Working Group, SLD Electroweak Group, SLD Heavy Flavour Group}),
  \bibinfo{journal}{Phys. Rept.} \textbf{\bibinfo{volume}{427}},
  \bibinfo{pages}{257} (\bibinfo{year}{2006}), \eprint{hep-ex/0509008}.

\bibitem[{\citenamefont{Baak et~al.}(2012{\natexlab{a}})\citenamefont{Baak,
  Goebel, Haller, Hoecker, Kennedy, Kogler, Moenig, Schott, and
  Stelzer}}]{Baak:2012kk}
\bibinfo{author}{\bibfnamefont{M.}~\bibnamefont{Baak}},
  \bibinfo{author}{\bibfnamefont{M.}~\bibnamefont{Goebel}},
  \bibinfo{author}{\bibfnamefont{J.}~\bibnamefont{Haller}},
  \bibinfo{author}{\bibfnamefont{A.}~\bibnamefont{Hoecker}},
  \bibinfo{author}{\bibfnamefont{D.}~\bibnamefont{Kennedy}},
  \bibinfo{author}{\bibfnamefont{R.}~\bibnamefont{Kogler}},
  \bibinfo{author}{\bibfnamefont{K.}~\bibnamefont{Moenig}},
  \bibinfo{author}{\bibfnamefont{M.}~\bibnamefont{Schott}}, \bibnamefont{and}
  \bibinfo{author}{\bibfnamefont{J.}~\bibnamefont{Stelzer}},
  \bibinfo{journal}{Eur. Phys. J. C} \textbf{\bibinfo{volume}{72}},
  \bibinfo{pages}{2205} (\bibinfo{year}{2012}{\natexlab{a}}),
  \eprint{1209.2716}.

\bibitem[{\citenamefont{Baak et~al.}(2014)\citenamefont{Baak, C\'uth, Haller,
  Hoecker, Kogler, M\"onig, Schott, and Stelzer}}]{Baak:2014ora}
\bibinfo{author}{\bibfnamefont{M.}~\bibnamefont{Baak}},
  \bibinfo{author}{\bibfnamefont{J.}~\bibnamefont{C\'uth}},
  \bibinfo{author}{\bibfnamefont{J.}~\bibnamefont{Haller}},
  \bibinfo{author}{\bibfnamefont{A.}~\bibnamefont{Hoecker}},
  \bibinfo{author}{\bibfnamefont{R.}~\bibnamefont{Kogler}},
  \bibinfo{author}{\bibfnamefont{K.}~\bibnamefont{M\"onig}},
  \bibinfo{author}{\bibfnamefont{M.}~\bibnamefont{Schott}}, \bibnamefont{and}
  \bibinfo{author}{\bibfnamefont{J.}~\bibnamefont{Stelzer}}
  (\bibinfo{collaboration}{Gfitter Group}), \bibinfo{journal}{Eur. Phys. J. C}
  \textbf{\bibinfo{volume}{74}}, \bibinfo{pages}{3046} (\bibinfo{year}{2014}),
  \eprint{1407.3792}.

\bibitem[{\citenamefont{Erler and Schott}(2019)}]{Erler:2019hds}
\bibinfo{author}{\bibfnamefont{J.}~\bibnamefont{Erler}} \bibnamefont{and}
  \bibinfo{author}{\bibfnamefont{M.}~\bibnamefont{Schott}},
  \bibinfo{journal}{Prog. Part. Nucl. Phys.} \textbf{\bibinfo{volume}{106}},
  \bibinfo{pages}{68} (\bibinfo{year}{2019}), \eprint{1902.05142}.

\bibitem[{\citenamefont{Patrignani et~al.}(2016)}]{ParticleDataGroup:2016lqr}
\bibinfo{author}{\bibfnamefont{C.}~\bibnamefont{Patrignani}}
  \bibnamefont{et~al.} (\bibinfo{collaboration}{Particle Data Group}),
  \bibinfo{journal}{Chin. Phys. C} \textbf{\bibinfo{volume}{40}},
  \bibinfo{pages}{100001} (\bibinfo{year}{2016}).

\bibitem[{\citenamefont{Aaltonen et~al.}(2022)}]{CDF:2022hxs}
\bibinfo{author}{\bibfnamefont{T.}~\bibnamefont{Aaltonen}} \bibnamefont{et~al.}
  (\bibinfo{collaboration}{CDF}), \bibinfo{journal}{Science}
  \textbf{\bibinfo{volume}{376}}, \bibinfo{pages}{170} (\bibinfo{year}{2022}).

\bibitem[{\citenamefont{Sirlin}(1980)}]{Sirlin:1980nh}
\bibinfo{author}{\bibfnamefont{A.}~\bibnamefont{Sirlin}},
  \bibinfo{journal}{Phys. Rev. D} \textbf{\bibinfo{volume}{22}},
  \bibinfo{pages}{971} (\bibinfo{year}{1980}).

\bibitem[{\citenamefont{Marciano and Sirlin}(1980)}]{Marciano:1980pb}
\bibinfo{author}{\bibfnamefont{W.~J.} \bibnamefont{Marciano}} \bibnamefont{and}
  \bibinfo{author}{\bibfnamefont{A.}~\bibnamefont{Sirlin}},
  \bibinfo{journal}{Phys. Rev. D} \textbf{\bibinfo{volume}{22}},
  \bibinfo{pages}{2695} (\bibinfo{year}{1980}), \bibinfo{note}{[Erratum:
  Phys.Rev.D 31, 213 (1985)]}.

\bibitem[{\citenamefont{Djouadi and Verzegnassi}(1987)}]{Djouadi:1987gn}
\bibinfo{author}{\bibfnamefont{A.}~\bibnamefont{Djouadi}} \bibnamefont{and}
  \bibinfo{author}{\bibfnamefont{C.}~\bibnamefont{Verzegnassi}},
  \bibinfo{journal}{Phys. Lett. B} \textbf{\bibinfo{volume}{195}},
  \bibinfo{pages}{265} (\bibinfo{year}{1987}).

\bibitem[{\citenamefont{Djouadi}(1988)}]{Djouadi:1987di}
\bibinfo{author}{\bibfnamefont{A.}~\bibnamefont{Djouadi}},
  \bibinfo{journal}{Nuovo Cim. A} \textbf{\bibinfo{volume}{100}},
  \bibinfo{pages}{357} (\bibinfo{year}{1988}).

\bibitem[{\citenamefont{Kniehl}(1990)}]{Kniehl:1989yc}
\bibinfo{author}{\bibfnamefont{B.~A.} \bibnamefont{Kniehl}},
  \bibinfo{journal}{Nucl. Phys. B} \textbf{\bibinfo{volume}{347}},
  \bibinfo{pages}{86} (\bibinfo{year}{1990}).

\bibitem[{\citenamefont{Halzen and Kniehl}(1991)}]{Halzen:1990je}
\bibinfo{author}{\bibfnamefont{F.}~\bibnamefont{Halzen}} \bibnamefont{and}
  \bibinfo{author}{\bibfnamefont{B.~A.} \bibnamefont{Kniehl}},
  \bibinfo{journal}{Nucl. Phys. B} \textbf{\bibinfo{volume}{353}},
  \bibinfo{pages}{567} (\bibinfo{year}{1991}).

\bibitem[{\citenamefont{Kniehl and Sirlin}(1992)}]{Kniehl:1991gu}
\bibinfo{author}{\bibfnamefont{B.~A.} \bibnamefont{Kniehl}} \bibnamefont{and}
  \bibinfo{author}{\bibfnamefont{A.}~\bibnamefont{Sirlin}},
  \bibinfo{journal}{Nucl. Phys. B} \textbf{\bibinfo{volume}{371}},
  \bibinfo{pages}{141} (\bibinfo{year}{1992}).

\bibitem[{\citenamefont{Freitas et~al.}(2000)\citenamefont{Freitas, Hollik,
  Walter, and Weiglein}}]{Freitas:2000gg}
\bibinfo{author}{\bibfnamefont{A.}~\bibnamefont{Freitas}},
  \bibinfo{author}{\bibfnamefont{W.}~\bibnamefont{Hollik}},
  \bibinfo{author}{\bibfnamefont{W.}~\bibnamefont{Walter}}, \bibnamefont{and}
  \bibinfo{author}{\bibfnamefont{G.}~\bibnamefont{Weiglein}},
  \bibinfo{journal}{Phys. Lett. B} \textbf{\bibinfo{volume}{495}},
  \bibinfo{pages}{338} (\bibinfo{year}{2000}), \bibinfo{note}{[Erratum:
  Phys.Lett.B 570, 265 (2003)]}, \eprint{hep-ph/0007091}.

\bibitem[{\citenamefont{Freitas et~al.}(2002)\citenamefont{Freitas, Hollik,
  Walter, and Weiglein}}]{Freitas:2002ja}
\bibinfo{author}{\bibfnamefont{A.}~\bibnamefont{Freitas}},
  \bibinfo{author}{\bibfnamefont{W.}~\bibnamefont{Hollik}},
  \bibinfo{author}{\bibfnamefont{W.}~\bibnamefont{Walter}}, \bibnamefont{and}
  \bibinfo{author}{\bibfnamefont{G.}~\bibnamefont{Weiglein}},
  \bibinfo{journal}{Nucl. Phys. B} \textbf{\bibinfo{volume}{632}},
  \bibinfo{pages}{189} (\bibinfo{year}{2002}), \bibinfo{note}{[Erratum:
  Nucl.Phys.B 666, 305--307 (2003)]}, \eprint{hep-ph/0202131}.

\bibitem[{\citenamefont{Awramik and Czakon}(2002)}]{Awramik:2002wn}
\bibinfo{author}{\bibfnamefont{M.}~\bibnamefont{Awramik}} \bibnamefont{and}
  \bibinfo{author}{\bibfnamefont{M.}~\bibnamefont{Czakon}},
  \bibinfo{journal}{Phys. Rev. Lett.} \textbf{\bibinfo{volume}{89}},
  \bibinfo{pages}{241801} (\bibinfo{year}{2002}), \eprint{hep-ph/0208113}.

\bibitem[{\citenamefont{Awramik and Czakon}(2003)}]{Awramik:2003ee}
\bibinfo{author}{\bibfnamefont{M.}~\bibnamefont{Awramik}} \bibnamefont{and}
  \bibinfo{author}{\bibfnamefont{M.}~\bibnamefont{Czakon}},
  \bibinfo{journal}{Phys. Lett. B} \textbf{\bibinfo{volume}{568}},
  \bibinfo{pages}{48} (\bibinfo{year}{2003}), \eprint{hep-ph/0305248}.

\bibitem[{\citenamefont{Onishchenko and Veretin}(2003)}]{Onishchenko:2002ve}
\bibinfo{author}{\bibfnamefont{A.}~\bibnamefont{Onishchenko}} \bibnamefont{and}
  \bibinfo{author}{\bibfnamefont{O.}~\bibnamefont{Veretin}},
  \bibinfo{journal}{Phys. Lett. B} \textbf{\bibinfo{volume}{551}},
  \bibinfo{pages}{111} (\bibinfo{year}{2003}), \eprint{hep-ph/0209010}.

\bibitem[{\citenamefont{Awramik et~al.}(2003)\citenamefont{Awramik, Czakon,
  Onishchenko, and Veretin}}]{Awramik:2002vu}
\bibinfo{author}{\bibfnamefont{M.}~\bibnamefont{Awramik}},
  \bibinfo{author}{\bibfnamefont{M.}~\bibnamefont{Czakon}},
  \bibinfo{author}{\bibfnamefont{A.}~\bibnamefont{Onishchenko}},
  \bibnamefont{and} \bibinfo{author}{\bibfnamefont{O.}~\bibnamefont{Veretin}},
  \bibinfo{journal}{Phys. Rev. D} \textbf{\bibinfo{volume}{68}},
  \bibinfo{pages}{053004} (\bibinfo{year}{2003}), \eprint{hep-ph/0209084}.

\bibitem[{\citenamefont{Avdeev et~al.}(1994)\citenamefont{Avdeev, Fleischer,
  Mikhailov, and Tarasov}}]{Avdeev:1994db}
\bibinfo{author}{\bibfnamefont{L.}~\bibnamefont{Avdeev}},
  \bibinfo{author}{\bibfnamefont{J.}~\bibnamefont{Fleischer}},
  \bibinfo{author}{\bibfnamefont{S.}~\bibnamefont{Mikhailov}},
  \bibnamefont{and} \bibinfo{author}{\bibfnamefont{O.}~\bibnamefont{Tarasov}},
  \bibinfo{journal}{Phys. Lett. B} \textbf{\bibinfo{volume}{336}},
  \bibinfo{pages}{560} (\bibinfo{year}{1994}), \bibinfo{note}{[Erratum:
  Phys.Lett.B 349, 597--598 (1995)]}, \eprint{hep-ph/9406363}.

\bibitem[{\citenamefont{Chetyrkin
  et~al.}(1995{\natexlab{a}})\citenamefont{Chetyrkin, Kuhn, and
  Steinhauser}}]{Chetyrkin:1995ix}
\bibinfo{author}{\bibfnamefont{K.~G.} \bibnamefont{Chetyrkin}},
  \bibinfo{author}{\bibfnamefont{J.~H.} \bibnamefont{Kuhn}}, \bibnamefont{and}
  \bibinfo{author}{\bibfnamefont{M.}~\bibnamefont{Steinhauser}},
  \bibinfo{journal}{Phys. Lett. B} \textbf{\bibinfo{volume}{351}},
  \bibinfo{pages}{331} (\bibinfo{year}{1995}{\natexlab{a}}),
  \eprint{hep-ph/9502291}.

\bibitem[{\citenamefont{Chetyrkin
  et~al.}(1995{\natexlab{b}})\citenamefont{Chetyrkin, Kuhn, and
  Steinhauser}}]{Chetyrkin:1995js}
\bibinfo{author}{\bibfnamefont{K.~G.} \bibnamefont{Chetyrkin}},
  \bibinfo{author}{\bibfnamefont{J.~H.} \bibnamefont{Kuhn}}, \bibnamefont{and}
  \bibinfo{author}{\bibfnamefont{M.}~\bibnamefont{Steinhauser}},
  \bibinfo{journal}{Phys. Rev. Lett.} \textbf{\bibinfo{volume}{75}},
  \bibinfo{pages}{3394} (\bibinfo{year}{1995}{\natexlab{b}}),
  \eprint{hep-ph/9504413}.

\bibitem[{\citenamefont{Chetyrkin et~al.}(1996)\citenamefont{Chetyrkin, Kuhn,
  and Steinhauser}}]{Chetyrkin:1996cf}
\bibinfo{author}{\bibfnamefont{K.~G.} \bibnamefont{Chetyrkin}},
  \bibinfo{author}{\bibfnamefont{J.~H.} \bibnamefont{Kuhn}}, \bibnamefont{and}
  \bibinfo{author}{\bibfnamefont{M.}~\bibnamefont{Steinhauser}},
  \bibinfo{journal}{Nucl. Phys. B} \textbf{\bibinfo{volume}{482}},
  \bibinfo{pages}{213} (\bibinfo{year}{1996}), \eprint{hep-ph/9606230}.

\bibitem[{\citenamefont{Faisst et~al.}(2003)\citenamefont{Faisst, Kuhn,
  Seidensticker, and Veretin}}]{Faisst:2003px}
\bibinfo{author}{\bibfnamefont{M.}~\bibnamefont{Faisst}},
  \bibinfo{author}{\bibfnamefont{J.~H.} \bibnamefont{Kuhn}},
  \bibinfo{author}{\bibfnamefont{T.}~\bibnamefont{Seidensticker}},
  \bibnamefont{and} \bibinfo{author}{\bibfnamefont{O.}~\bibnamefont{Veretin}},
  \bibinfo{journal}{Nucl. Phys. B} \textbf{\bibinfo{volume}{665}},
  \bibinfo{pages}{649} (\bibinfo{year}{2003}), \eprint{hep-ph/0302275}.

\bibitem[{\citenamefont{van~der Bij et~al.}(2001)\citenamefont{van~der Bij,
  Chetyrkin, Faisst, Jikia, and Seidensticker}}]{vanderBij:2000cg}
\bibinfo{author}{\bibfnamefont{J.~J.} \bibnamefont{van~der Bij}},
  \bibinfo{author}{\bibfnamefont{K.~G.} \bibnamefont{Chetyrkin}},
  \bibinfo{author}{\bibfnamefont{M.}~\bibnamefont{Faisst}},
  \bibinfo{author}{\bibfnamefont{G.}~\bibnamefont{Jikia}}, \bibnamefont{and}
  \bibinfo{author}{\bibfnamefont{T.}~\bibnamefont{Seidensticker}},
  \bibinfo{journal}{Phys. Lett. B} \textbf{\bibinfo{volume}{498}},
  \bibinfo{pages}{156} (\bibinfo{year}{2001}), \eprint{hep-ph/0011373}.

\bibitem[{\citenamefont{Boughezal et~al.}(2005)\citenamefont{Boughezal, Tausk,
  and van~der Bij}}]{Boughezal:2004ef}
\bibinfo{author}{\bibfnamefont{R.}~\bibnamefont{Boughezal}},
  \bibinfo{author}{\bibfnamefont{J.~B.} \bibnamefont{Tausk}}, \bibnamefont{and}
  \bibinfo{author}{\bibfnamefont{J.~J.} \bibnamefont{van~der Bij}},
  \bibinfo{journal}{Nucl. Phys. B} \textbf{\bibinfo{volume}{713}},
  \bibinfo{pages}{278} (\bibinfo{year}{2005}), \eprint{hep-ph/0410216}.

\bibitem[{\citenamefont{Boughezal and Czakon}(2006)}]{Boughezal:2006xk}
\bibinfo{author}{\bibfnamefont{R.}~\bibnamefont{Boughezal}} \bibnamefont{and}
  \bibinfo{author}{\bibfnamefont{M.}~\bibnamefont{Czakon}},
  \bibinfo{journal}{Nucl. Phys. B} \textbf{\bibinfo{volume}{755}},
  \bibinfo{pages}{221} (\bibinfo{year}{2006}), \eprint{hep-ph/0606232}.

\bibitem[{\citenamefont{Chetyrkin et~al.}(2006)\citenamefont{Chetyrkin, Faisst,
  Kuhn, Maierhofer, and Sturm}}]{Chetyrkin:2006bj}
\bibinfo{author}{\bibfnamefont{K.~G.} \bibnamefont{Chetyrkin}},
  \bibinfo{author}{\bibfnamefont{M.}~\bibnamefont{Faisst}},
  \bibinfo{author}{\bibfnamefont{J.~H.} \bibnamefont{Kuhn}},
  \bibinfo{author}{\bibfnamefont{P.}~\bibnamefont{Maierhofer}},
  \bibnamefont{and} \bibinfo{author}{\bibfnamefont{C.}~\bibnamefont{Sturm}},
  \bibinfo{journal}{Phys. Rev. Lett.} \textbf{\bibinfo{volume}{97}},
  \bibinfo{pages}{102003} (\bibinfo{year}{2006}), \eprint{hep-ph/0605201}.

\bibitem[{\citenamefont{Schroder and Steinhauser}(2005)}]{Schroder:2005db}
\bibinfo{author}{\bibfnamefont{Y.}~\bibnamefont{Schroder}} \bibnamefont{and}
  \bibinfo{author}{\bibfnamefont{M.}~\bibnamefont{Steinhauser}},
  \bibinfo{journal}{Phys. Lett. B} \textbf{\bibinfo{volume}{622}},
  \bibinfo{pages}{124} (\bibinfo{year}{2005}), \eprint{hep-ph/0504055}.

\bibitem[{\citenamefont{Awramik et~al.}(2006)\citenamefont{Awramik, Czakon, and
  Freitas}}]{Awramik:2006uz}
\bibinfo{author}{\bibfnamefont{M.}~\bibnamefont{Awramik}},
  \bibinfo{author}{\bibfnamefont{M.}~\bibnamefont{Czakon}}, \bibnamefont{and}
  \bibinfo{author}{\bibfnamefont{A.}~\bibnamefont{Freitas}},
  \bibinfo{journal}{JHEP} \textbf{\bibinfo{volume}{11}}, \bibinfo{pages}{048}
  (\bibinfo{year}{2006}), \eprint{hep-ph/0608099}.

\bibitem[{\citenamefont{Degrassi et~al.}(2015)\citenamefont{Degrassi, Gambino,
  and Giardino}}]{Degrassi:2014sxa}
\bibinfo{author}{\bibfnamefont{G.}~\bibnamefont{Degrassi}},
  \bibinfo{author}{\bibfnamefont{P.}~\bibnamefont{Gambino}}, \bibnamefont{and}
  \bibinfo{author}{\bibfnamefont{P.~P.} \bibnamefont{Giardino}},
  \bibinfo{journal}{JHEP} \textbf{\bibinfo{volume}{05}}, \bibinfo{pages}{154}
  (\bibinfo{year}{2015}), \eprint{1411.7040}.

\bibitem[{\citenamefont{Awramik et~al.}(2004)\citenamefont{Awramik, Czakon,
  Freitas, and Weiglein}}]{Awramik:2003rn}
\bibinfo{author}{\bibfnamefont{M.}~\bibnamefont{Awramik}},
  \bibinfo{author}{\bibfnamefont{M.}~\bibnamefont{Czakon}},
  \bibinfo{author}{\bibfnamefont{A.}~\bibnamefont{Freitas}}, \bibnamefont{and}
  \bibinfo{author}{\bibfnamefont{G.}~\bibnamefont{Weiglein}},
  \bibinfo{journal}{Phys. Rev. D} \textbf{\bibinfo{volume}{69}},
  \bibinfo{pages}{053006} (\bibinfo{year}{2004}), \eprint{hep-ph/0311148}.

\bibitem[{\citenamefont{Zyla et~al.}(2020)}]{ParticleDataGroup:2020ssz}
\bibinfo{author}{\bibfnamefont{P.~A.} \bibnamefont{Zyla}} \bibnamefont{et~al.}
  (\bibinfo{collaboration}{Particle Data Group}), \bibinfo{journal}{PTEP}
  \textbf{\bibinfo{volume}{2020}}, \bibinfo{pages}{083C01}
  (\bibinfo{year}{2020}).

\bibitem[{\citenamefont{Crivellin et~al.}(2020)\citenamefont{Crivellin,
  Hoferichter, Manzari, and Montull}}]{Crivellin:2020zul}
\bibinfo{author}{\bibfnamefont{A.}~\bibnamefont{Crivellin}},
  \bibinfo{author}{\bibfnamefont{M.}~\bibnamefont{Hoferichter}},
  \bibinfo{author}{\bibfnamefont{C.~A.} \bibnamefont{Manzari}},
  \bibnamefont{and} \bibinfo{author}{\bibfnamefont{M.}~\bibnamefont{Montull}},
  \bibinfo{journal}{Phys. Rev. Lett.} \textbf{\bibinfo{volume}{125}},
  \bibinfo{pages}{091801} (\bibinfo{year}{2020}), \eprint{2003.04886}.

\bibitem[{\citenamefont{Davier et~al.}(2020)\citenamefont{Davier, Hoecker,
  Malaescu, and Zhang}}]{Davier:2019can}
\bibinfo{author}{\bibfnamefont{M.}~\bibnamefont{Davier}},
  \bibinfo{author}{\bibfnamefont{A.}~\bibnamefont{Hoecker}},
  \bibinfo{author}{\bibfnamefont{B.}~\bibnamefont{Malaescu}}, \bibnamefont{and}
  \bibinfo{author}{\bibfnamefont{Z.}~\bibnamefont{Zhang}},
  \bibinfo{journal}{Eur. Phys. J. C} \textbf{\bibinfo{volume}{80}},
  \bibinfo{pages}{241} (\bibinfo{year}{2020}), \bibinfo{note}{[Erratum:
  Eur.Phys.J.C 80, 410 (2020)]}, \eprint{1908.00921}.

\bibitem[{\citenamefont{Keshavarzi
  et~al.}(2020{\natexlab{a}})\citenamefont{Keshavarzi, Nomura, and
  Teubner}}]{Keshavarzi:2019abf}
\bibinfo{author}{\bibfnamefont{A.}~\bibnamefont{Keshavarzi}},
  \bibinfo{author}{\bibfnamefont{D.}~\bibnamefont{Nomura}}, \bibnamefont{and}
  \bibinfo{author}{\bibfnamefont{T.}~\bibnamefont{Teubner}},
  \bibinfo{journal}{Phys. Rev. D} \textbf{\bibinfo{volume}{101}},
  \bibinfo{pages}{014029} (\bibinfo{year}{2020}{\natexlab{a}}),
  \eprint{1911.00367}.

\bibitem[{\citenamefont{Aaltonen et~al.}(2018)}]{CDF:2018cnj}
\bibinfo{author}{\bibfnamefont{T.~A.} \bibnamefont{Aaltonen}}
  \bibnamefont{et~al.} (\bibinfo{collaboration}{CDF, D0}),
  \bibinfo{journal}{Phys. Rev. D} \textbf{\bibinfo{volume}{97}},
  \bibinfo{pages}{112007} (\bibinfo{year}{2018}), \eprint{1801.06283}.

\bibitem[{\citenamefont{Flacher et~al.}(2009)\citenamefont{Flacher, Goebel,
  Haller, Hocker, Monig, and Stelzer}}]{Flacher:2008zq}
\bibinfo{author}{\bibfnamefont{H.}~\bibnamefont{Flacher}},
  \bibinfo{author}{\bibfnamefont{M.}~\bibnamefont{Goebel}},
  \bibinfo{author}{\bibfnamefont{J.}~\bibnamefont{Haller}},
  \bibinfo{author}{\bibfnamefont{A.}~\bibnamefont{Hocker}},
  \bibinfo{author}{\bibfnamefont{K.}~\bibnamefont{Monig}}, \bibnamefont{and}
  \bibinfo{author}{\bibfnamefont{J.}~\bibnamefont{Stelzer}},
  \bibinfo{journal}{Eur. Phys. J. C} \textbf{\bibinfo{volume}{60}},
  \bibinfo{pages}{543} (\bibinfo{year}{2009}), \bibinfo{note}{[Erratum:
  Eur.Phys.J.C 71, 1718 (2011)]}, \eprint{0811.0009}.

\bibitem[{\citenamefont{Baak et~al.}(2012{\natexlab{b}})\citenamefont{Baak,
  Goebel, Haller, Hoecker, Ludwig, Moenig, Schott, and Stelzer}}]{Baak:2011ze}
\bibinfo{author}{\bibfnamefont{M.}~\bibnamefont{Baak}},
  \bibinfo{author}{\bibfnamefont{M.}~\bibnamefont{Goebel}},
  \bibinfo{author}{\bibfnamefont{J.}~\bibnamefont{Haller}},
  \bibinfo{author}{\bibfnamefont{A.}~\bibnamefont{Hoecker}},
  \bibinfo{author}{\bibfnamefont{D.}~\bibnamefont{Ludwig}},
  \bibinfo{author}{\bibfnamefont{K.}~\bibnamefont{Moenig}},
  \bibinfo{author}{\bibfnamefont{M.}~\bibnamefont{Schott}}, \bibnamefont{and}
  \bibinfo{author}{\bibfnamefont{J.}~\bibnamefont{Stelzer}},
  \bibinfo{journal}{Eur. Phys. J. C} \textbf{\bibinfo{volume}{72}},
  \bibinfo{pages}{2003} (\bibinfo{year}{2012}{\natexlab{b}}),
  \eprint{1107.0975}.

\bibitem[{\citenamefont{Haller et~al.}(2018)\citenamefont{Haller, Hoecker,
  Kogler, M\"onig, Peiffer, and Stelzer}}]{Haller:2018nnx}
\bibinfo{author}{\bibfnamefont{J.}~\bibnamefont{Haller}},
  \bibinfo{author}{\bibfnamefont{A.}~\bibnamefont{Hoecker}},
  \bibinfo{author}{\bibfnamefont{R.}~\bibnamefont{Kogler}},
  \bibinfo{author}{\bibfnamefont{K.}~\bibnamefont{M\"onig}},
  \bibinfo{author}{\bibfnamefont{T.}~\bibnamefont{Peiffer}}, \bibnamefont{and}
  \bibinfo{author}{\bibfnamefont{J.}~\bibnamefont{Stelzer}},
  \bibinfo{journal}{Eur. Phys. J. C} \textbf{\bibinfo{volume}{78}},
  \bibinfo{pages}{675} (\bibinfo{year}{2018}), \eprint{1803.01853}.

\bibitem[{\citenamefont{De~Blas et~al.}(2020)}]{DeBlas:2019ehy}
\bibinfo{author}{\bibfnamefont{J.}~\bibnamefont{De~Blas}} \bibnamefont{et~al.},
  \bibinfo{journal}{Eur. Phys. J. C} \textbf{\bibinfo{volume}{80}},
  \bibinfo{pages}{456} (\bibinfo{year}{2020}), \eprint{1910.14012}.

\bibitem[{\citenamefont{Abazov et~al.}(2014)}]{D0:2014yem}
\bibinfo{author}{\bibfnamefont{V.~M.} \bibnamefont{Abazov}}
  \bibnamefont{et~al.} (\bibinfo{collaboration}{D0}), \bibinfo{journal}{Phys.
  Rev. Lett.} \textbf{\bibinfo{volume}{113}}, \bibinfo{pages}{032002}
  (\bibinfo{year}{2014}), \eprint{1405.1756}.

\bibitem[{\citenamefont{Sirunyan et~al.}(2018)}]{CMS:2018quc}
\bibinfo{author}{\bibfnamefont{A.~M.} \bibnamefont{Sirunyan}}
  \bibnamefont{et~al.} (\bibinfo{collaboration}{CMS}), \bibinfo{journal}{Eur.
  Phys. J. C} \textbf{\bibinfo{volume}{78}}, \bibinfo{pages}{891}
  (\bibinfo{year}{2018}), \eprint{1805.01428}.

\bibitem[{\citenamefont{Keshavarzi
  et~al.}(2020{\natexlab{b}})\citenamefont{Keshavarzi, Marciano, Passera, and
  Sirlin}}]{Keshavarzi:2020bfy}
\bibinfo{author}{\bibfnamefont{A.}~\bibnamefont{Keshavarzi}},
  \bibinfo{author}{\bibfnamefont{W.~J.} \bibnamefont{Marciano}},
  \bibinfo{author}{\bibfnamefont{M.}~\bibnamefont{Passera}}, \bibnamefont{and}
  \bibinfo{author}{\bibfnamefont{A.}~\bibnamefont{Sirlin}},
  \bibinfo{journal}{Phys. Rev. D} \textbf{\bibinfo{volume}{102}},
  \bibinfo{pages}{033002} (\bibinfo{year}{2020}{\natexlab{b}}),
  \eprint{2006.12666}.

\bibitem[{\citenamefont{de~Rafael}(2020)}]{deRafael:2020uif}
\bibinfo{author}{\bibfnamefont{E.}~\bibnamefont{de~Rafael}},
  \bibinfo{journal}{Phys. Rev. D} \textbf{\bibinfo{volume}{102}},
  \bibinfo{pages}{056025} (\bibinfo{year}{2020}), \eprint{2006.13880}.

\bibitem[{\citenamefont{Altarelli and Barbieri}(1991)}]{Altarelli:1990zd}
\bibinfo{author}{\bibfnamefont{G.}~\bibnamefont{Altarelli}} \bibnamefont{and}
  \bibinfo{author}{\bibfnamefont{R.}~\bibnamefont{Barbieri}},
  \bibinfo{journal}{Phys. Lett. B} \textbf{\bibinfo{volume}{253}},
  \bibinfo{pages}{161} (\bibinfo{year}{1991}).

\bibitem[{\citenamefont{Altarelli et~al.}(1992)\citenamefont{Altarelli,
  Barbieri, and Jadach}}]{Altarelli:1991fk}
\bibinfo{author}{\bibfnamefont{G.}~\bibnamefont{Altarelli}},
  \bibinfo{author}{\bibfnamefont{R.}~\bibnamefont{Barbieri}}, \bibnamefont{and}
  \bibinfo{author}{\bibfnamefont{S.}~\bibnamefont{Jadach}},
  \bibinfo{journal}{Nucl. Phys. B} \textbf{\bibinfo{volume}{369}},
  \bibinfo{pages}{3} (\bibinfo{year}{1992}), \bibinfo{note}{[Erratum:
  Nucl.Phys.B 376, 444 (1992)]}.

\bibitem[{\citenamefont{Barbieri et~al.}(1992)\citenamefont{Barbieri, Frigeni,
  and Caravaglios}}]{Barbieri:1991qp}
\bibinfo{author}{\bibfnamefont{R.}~\bibnamefont{Barbieri}},
  \bibinfo{author}{\bibfnamefont{M.}~\bibnamefont{Frigeni}}, \bibnamefont{and}
  \bibinfo{author}{\bibfnamefont{F.}~\bibnamefont{Caravaglios}},
  \bibinfo{journal}{Phys. Lett. B} \textbf{\bibinfo{volume}{279}},
  \bibinfo{pages}{169} (\bibinfo{year}{1992}).

\bibitem[{\citenamefont{Altarelli et~al.}(1993)\citenamefont{Altarelli,
  Barbieri, and Caravaglios}}]{Altarelli:1993bh}
\bibinfo{author}{\bibfnamefont{G.}~\bibnamefont{Altarelli}},
  \bibinfo{author}{\bibfnamefont{R.}~\bibnamefont{Barbieri}}, \bibnamefont{and}
  \bibinfo{author}{\bibfnamefont{F.}~\bibnamefont{Caravaglios}},
  \bibinfo{journal}{Phys. Lett. B} \textbf{\bibinfo{volume}{314}},
  \bibinfo{pages}{357} (\bibinfo{year}{1993}).

\bibitem[{\citenamefont{Burgess et~al.}(1994)\citenamefont{Burgess, Godfrey,
  Konig, London, and Maksymyk}}]{Burgess:1993vc}
\bibinfo{author}{\bibfnamefont{C.~P.} \bibnamefont{Burgess}},
  \bibinfo{author}{\bibfnamefont{S.}~\bibnamefont{Godfrey}},
  \bibinfo{author}{\bibfnamefont{H.}~\bibnamefont{Konig}},
  \bibinfo{author}{\bibfnamefont{D.}~\bibnamefont{London}}, \bibnamefont{and}
  \bibinfo{author}{\bibfnamefont{I.}~\bibnamefont{Maksymyk}},
  \bibinfo{journal}{Phys. Rev. D} \textbf{\bibinfo{volume}{49}},
  \bibinfo{pages}{6115} (\bibinfo{year}{1994}), \eprint{hep-ph/9312291}.

\bibitem[{\citenamefont{Peskin and Takeuchi}(1992)}]{Peskin:1991sw}
\bibinfo{author}{\bibfnamefont{M.~E.} \bibnamefont{Peskin}} \bibnamefont{and}
  \bibinfo{author}{\bibfnamefont{T.}~\bibnamefont{Takeuchi}},
  \bibinfo{journal}{Phys. Rev. D} \textbf{\bibinfo{volume}{46}},
  \bibinfo{pages}{381} (\bibinfo{year}{1992}).

\bibitem[{\citenamefont{Lavoura and Li}(1994)}]{Lavoura:1993nq}
\bibinfo{author}{\bibfnamefont{L.}~\bibnamefont{Lavoura}} \bibnamefont{and}
  \bibinfo{author}{\bibfnamefont{L.-F.} \bibnamefont{Li}},
  \bibinfo{journal}{Phys. Rev. D} \textbf{\bibinfo{volume}{49}},
  \bibinfo{pages}{1409} (\bibinfo{year}{1994}), \eprint{hep-ph/9309262}.

\bibitem[{\citenamefont{Lu et~al.}(2022)\citenamefont{Lu, Wu, Wu, and
  Zhu}}]{2hdmew}
\bibinfo{author}{\bibfnamefont{C.-T.} \bibnamefont{Lu}},
  \bibinfo{author}{\bibfnamefont{L.}~\bibnamefont{Wu}},
  \bibinfo{author}{\bibfnamefont{Y.}~\bibnamefont{Wu}}, \bibnamefont{and}
  \bibinfo{author}{\bibfnamefont{B.}~\bibnamefont{Zhu}} (\bibinfo{year}{2022}),
  \eprint{to appear}.

\bibitem[{\citenamefont{Branco et~al.}(2012)\citenamefont{Branco, Ferreira,
  Lavoura, Rebelo, Sher, and Silva}}]{Branco:2011iw}
\bibinfo{author}{\bibfnamefont{G.~C.} \bibnamefont{Branco}},
  \bibinfo{author}{\bibfnamefont{P.~M.} \bibnamefont{Ferreira}},
  \bibinfo{author}{\bibfnamefont{L.}~\bibnamefont{Lavoura}},
  \bibinfo{author}{\bibfnamefont{M.~N.} \bibnamefont{Rebelo}},
  \bibinfo{author}{\bibfnamefont{M.}~\bibnamefont{Sher}}, \bibnamefont{and}
  \bibinfo{author}{\bibfnamefont{J.~P.} \bibnamefont{Silva}},
  \bibinfo{journal}{Phys. Rept.} \textbf{\bibinfo{volume}{516}},
  \bibinfo{pages}{1} (\bibinfo{year}{2012}), \eprint{1106.0034}.

\bibitem[{\citenamefont{Gunion and Haber}(2003)}]{Gunion:2002zf}
\bibinfo{author}{\bibfnamefont{J.~F.} \bibnamefont{Gunion}} \bibnamefont{and}
  \bibinfo{author}{\bibfnamefont{H.~E.} \bibnamefont{Haber}},
  \bibinfo{journal}{Phys. Rev. D} \textbf{\bibinfo{volume}{67}},
  \bibinfo{pages}{075019} (\bibinfo{year}{2003}), \eprint{hep-ph/0207010}.

\bibitem[{\citenamefont{Craig et~al.}(2013)\citenamefont{Craig, Galloway, and
  Thomas}}]{Craig:2013hca}
\bibinfo{author}{\bibfnamefont{N.}~\bibnamefont{Craig}},
  \bibinfo{author}{\bibfnamefont{J.}~\bibnamefont{Galloway}}, \bibnamefont{and}
  \bibinfo{author}{\bibfnamefont{S.}~\bibnamefont{Thomas}}
  (\bibinfo{year}{2013}), \eprint{1305.2424}.

\bibitem[{\citenamefont{Carena et~al.}(2014)\citenamefont{Carena, Low, Shah,
  and Wagner}}]{Carena:2013ooa}
\bibinfo{author}{\bibfnamefont{M.}~\bibnamefont{Carena}},
  \bibinfo{author}{\bibfnamefont{I.}~\bibnamefont{Low}},
  \bibinfo{author}{\bibfnamefont{N.~R.} \bibnamefont{Shah}}, \bibnamefont{and}
  \bibinfo{author}{\bibfnamefont{C.~E.~M.} \bibnamefont{Wagner}},
  \bibinfo{journal}{JHEP} \textbf{\bibinfo{volume}{04}}, \bibinfo{pages}{015}
  (\bibinfo{year}{2014}), \eprint{1310.2248}.

\bibitem[{\citenamefont{Bernon et~al.}(2015)\citenamefont{Bernon, Gunion,
  Haber, Jiang, and Kraml}}]{Bernon:2015qea}
\bibinfo{author}{\bibfnamefont{J.}~\bibnamefont{Bernon}},
  \bibinfo{author}{\bibfnamefont{J.~F.} \bibnamefont{Gunion}},
  \bibinfo{author}{\bibfnamefont{H.~E.} \bibnamefont{Haber}},
  \bibinfo{author}{\bibfnamefont{Y.}~\bibnamefont{Jiang}}, \bibnamefont{and}
  \bibinfo{author}{\bibfnamefont{S.}~\bibnamefont{Kraml}},
  \bibinfo{journal}{Phys. Rev. D} \textbf{\bibinfo{volume}{92}},
  \bibinfo{pages}{075004} (\bibinfo{year}{2015}), \eprint{1507.00933}.

\bibitem[{\citenamefont{Chen et~al.}(2019)\citenamefont{Chen, Han, Su, Su, and
  Wu}}]{Chen:2018shg}
\bibinfo{author}{\bibfnamefont{N.}~\bibnamefont{Chen}},
  \bibinfo{author}{\bibfnamefont{T.}~\bibnamefont{Han}},
  \bibinfo{author}{\bibfnamefont{S.}~\bibnamefont{Su}},
  \bibinfo{author}{\bibfnamefont{W.}~\bibnamefont{Su}}, \bibnamefont{and}
  \bibinfo{author}{\bibfnamefont{Y.}~\bibnamefont{Wu}}, \bibinfo{journal}{JHEP}
  \textbf{\bibinfo{volume}{03}}, \bibinfo{pages}{023} (\bibinfo{year}{2019}),
  \eprint{1808.02037}.

\bibitem[{\citenamefont{Eberhardt et~al.}(2021)\citenamefont{Eberhardt,
  Mart\'\i{}nez, and Pich}}]{Eberhardt:2020dat}
\bibinfo{author}{\bibfnamefont{O.}~\bibnamefont{Eberhardt}},
  \bibinfo{author}{\bibfnamefont{A.~P.~n.} \bibnamefont{Mart\'\i{}nez}},
  \bibnamefont{and} \bibinfo{author}{\bibfnamefont{A.}~\bibnamefont{Pich}},
  \bibinfo{journal}{JHEP} \textbf{\bibinfo{volume}{05}}, \bibinfo{pages}{005}
  (\bibinfo{year}{2021}), \eprint{2012.09200}.

\bibitem[{\citenamefont{Aad et~al.}(2021)}]{ATLAS:2021upq}
\bibinfo{author}{\bibfnamefont{G.}~\bibnamefont{Aad}} \bibnamefont{et~al.}
  (\bibinfo{collaboration}{ATLAS}), \bibinfo{journal}{JHEP}
  \textbf{\bibinfo{volume}{06}}, \bibinfo{pages}{145} (\bibinfo{year}{2021}),
  \eprint{2102.10076}.

\bibitem[{\citenamefont{Sirunyan et~al.}(2020{\natexlab{a}})}]{CMS:2020imj}
\bibinfo{author}{\bibfnamefont{A.~M.} \bibnamefont{Sirunyan}}
  \bibnamefont{et~al.} (\bibinfo{collaboration}{CMS}), \bibinfo{journal}{JHEP}
  \textbf{\bibinfo{volume}{07}}, \bibinfo{pages}{126}
  (\bibinfo{year}{2020}{\natexlab{a}}), \eprint{2001.07763}.

\bibitem[{ATL(2022)}]{ATLAS-CONF-2022-008}
\bibinfo{type}{Tech. Rep.}, \bibinfo{institution}{CERN},
  \bibinfo{address}{Geneva} (\bibinfo{year}{2022}),
  \urlprefix\url{https://cds.cern.ch/record/2805212}.

\bibitem[{\citenamefont{Sirunyan et~al.}(2020{\natexlab{b}})}]{CMS:2019rvj}
\bibinfo{author}{\bibfnamefont{A.~M.} \bibnamefont{Sirunyan}}
  \bibnamefont{et~al.} (\bibinfo{collaboration}{CMS}), \bibinfo{journal}{Eur.
  Phys. J. C} \textbf{\bibinfo{volume}{80}}, \bibinfo{pages}{75}
  (\bibinfo{year}{2020}{\natexlab{b}}), \eprint{1908.06463}.

\bibitem[{\citenamefont{'t~Hooft and Veltman}(1979)}]{tHooft:1978jhc}
\bibinfo{author}{\bibfnamefont{G.}~\bibnamefont{'t~Hooft}} \bibnamefont{and}
  \bibinfo{author}{\bibfnamefont{M.~J.~G.} \bibnamefont{Veltman}},
  \bibinfo{journal}{Nucl. Phys. B} \textbf{\bibinfo{volume}{153}},
  \bibinfo{pages}{365} (\bibinfo{year}{1979}).

\bibitem[{\citenamefont{Hahn and Perez-Victoria}(1999)}]{Hahn:1998yk}
\bibinfo{author}{\bibfnamefont{T.}~\bibnamefont{Hahn}} \bibnamefont{and}
  \bibinfo{author}{\bibfnamefont{M.}~\bibnamefont{Perez-Victoria}},
  \bibinfo{journal}{Comput. Phys. Commun.} \textbf{\bibinfo{volume}{118}},
  \bibinfo{pages}{153} (\bibinfo{year}{1999}), \eprint{hep-ph/9807565}.

\end{thebibliography}

\onecolumngrid
\clearpage

\setcounter{page}{1}
\setcounter{equation}{0}
\setcounter{figure}{0}
\setcounter{table}{0}
\setcounter{section}{0}
\setcounter{subsection}{0}
\renewcommand{\theequation}{S.\arabic{equation}}
\renewcommand{\thefigure}{S\arabic{figure}}
\renewcommand{\thetable}{S\arabic{table}}
\renewcommand{\thesection}{\Roman{section}}
\renewcommand{\thesubsection}{\Alph{subsection}}
\newcommand{\ssection}[1]{
    \addtocounter{section}{1}
    \section{\thesection.~~~#1}
    \addtocounter{section}{-1}
    \refstepcounter{section}
}
\newcommand{\ssubsection}[1]{
    \addtocounter{subsection}{1}
    \subsection{\thesubsection.~~~#1}
    \addtocounter{subsection}{-1}
    \refstepcounter{subsection}
}
\newcommand{\fakeaffil}[2]{$^{#1}$\textit{#2}\\}

\thispagestyle{empty}
\begin{center}
    \begin{spacing}{1.2}
        \textbf{\large
            Supplemental Material:\\ Electroweak Precision Fit and New Physics in light of $W$ Boson Mass
        }
    \end{spacing}
    \par\smallskip
    Chih-Ting Lu,$^{1}$
    Lei Wu,$^{1}$
    Yongcheng Wu,$^{1}$
    and Bin Zhu,$^{2}$
    \par
    {\small
        \fakeaffil{1}{Department of Physics and Institute of Theoretical Physics, Nanjing Normal University, Nanjing, 210023, China}
        \fakeaffil{4}{Department of Physics, Yantai University, Yantai 264005, China}
    }

\end{center}
\par\smallskip

\begin{figure}[ht]
    \centering
    \includegraphics[height=15cm]{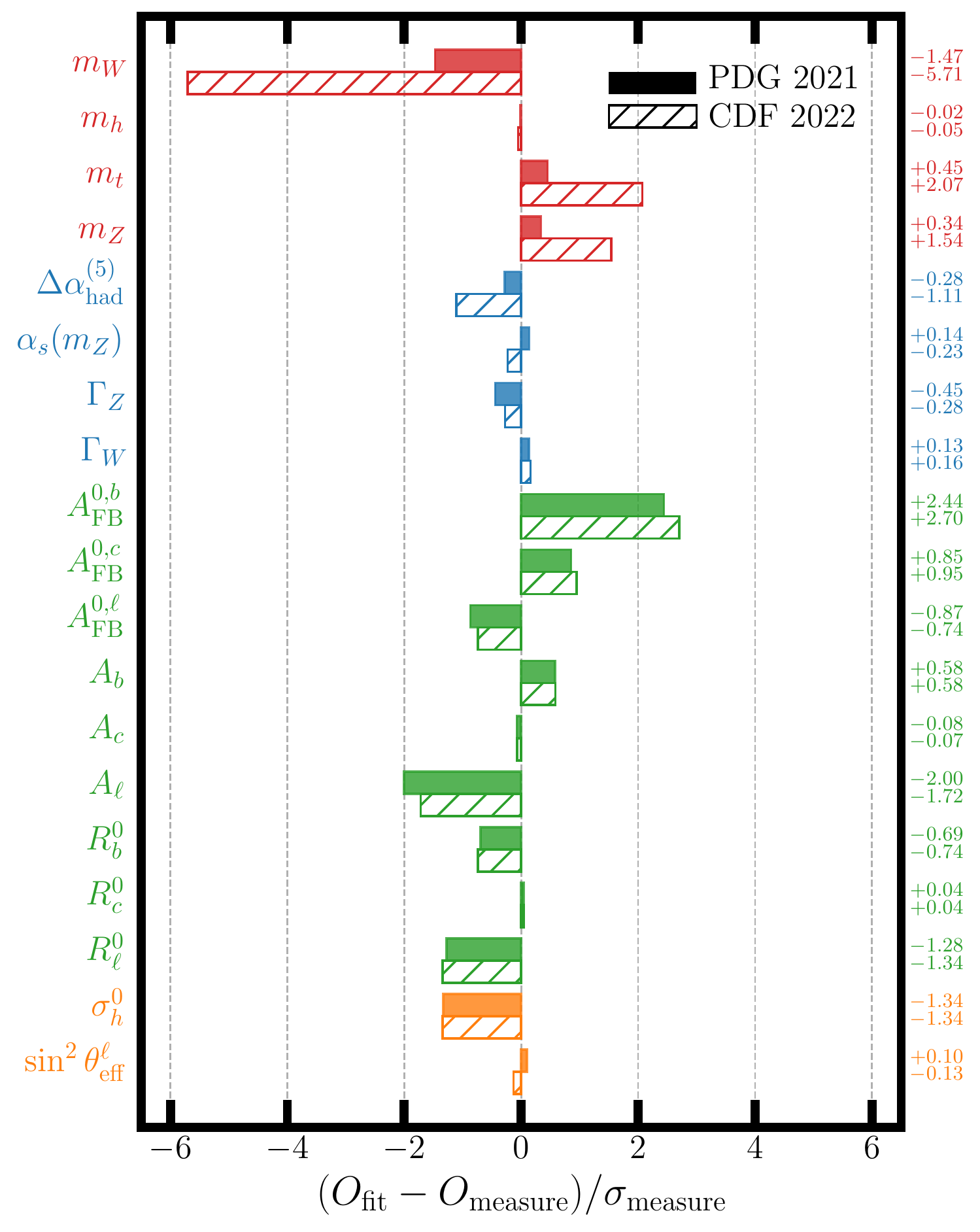}
    \caption{The comparison of the ``pull'' defined in Tab.~\ref{tab:fit} between EW fits using the PDG 2021 data set with the old value of $m_W$ and the new CDF value of $m_W$ in the SM.}
    \label{fig:pull}
\end{figure}

\begin{figure}[ht]

\includegraphics[width=8cm,height=6cm]{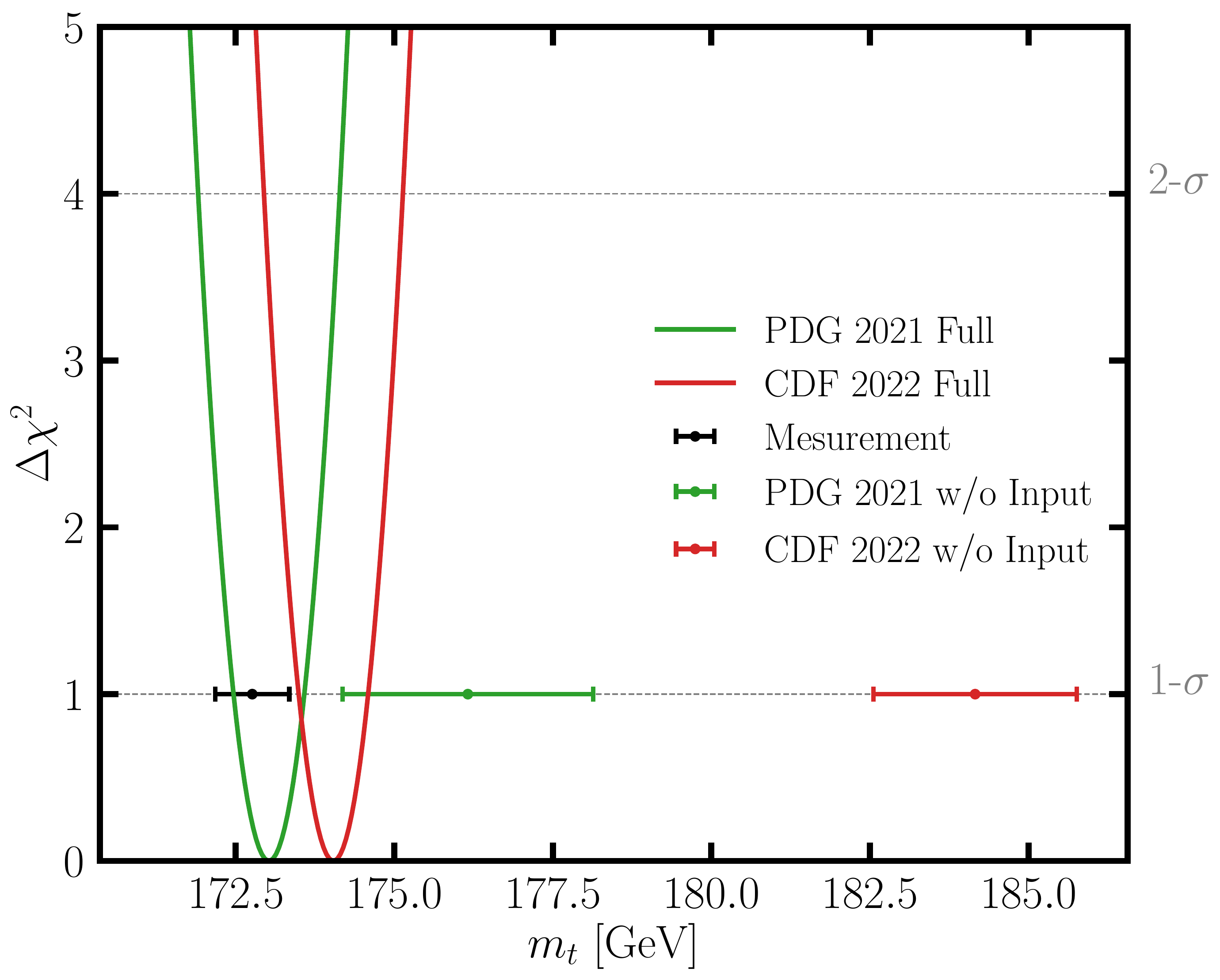}
\includegraphics[width=8cm,height=6cm]{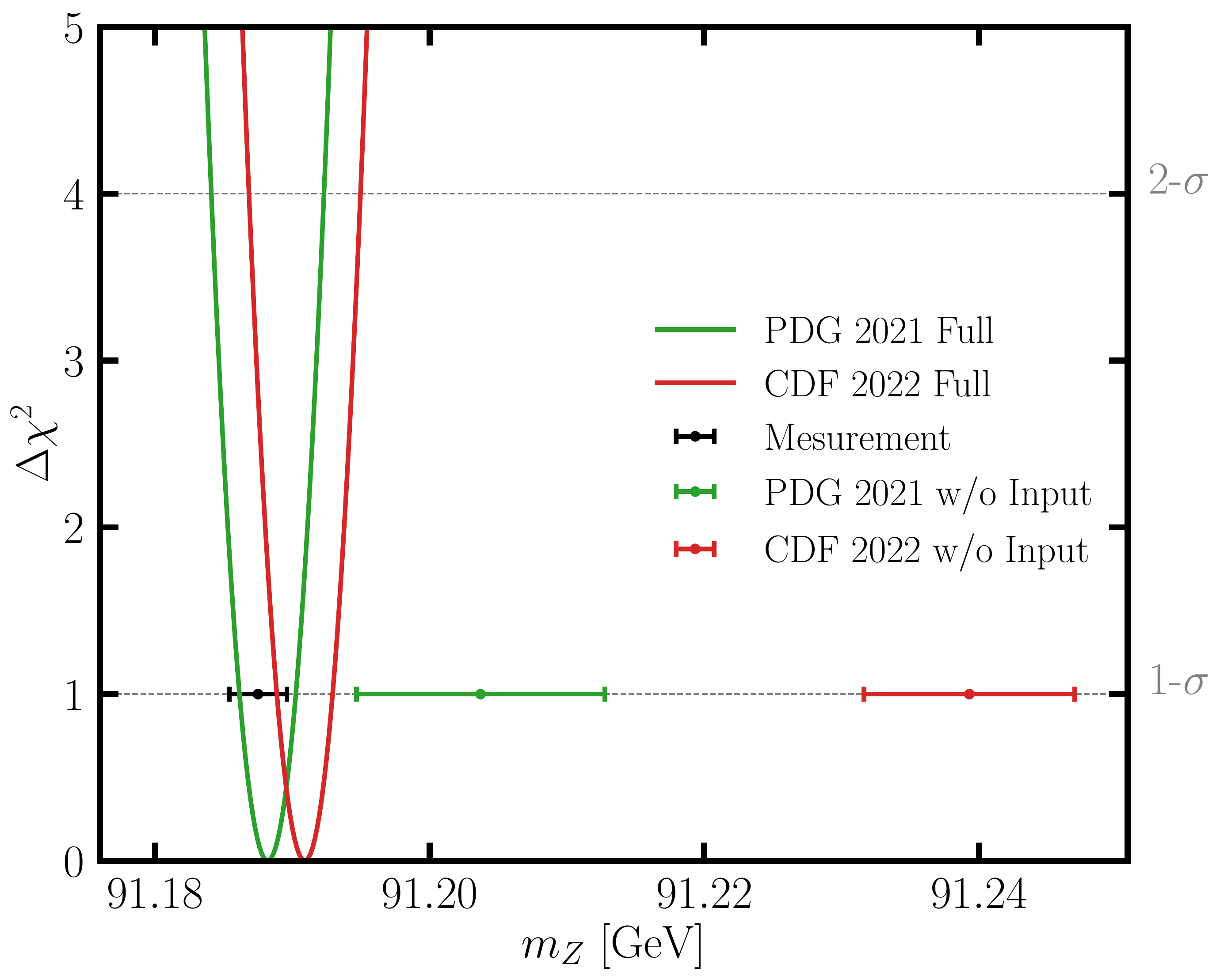}
\includegraphics[width=8cm,height=6cm]{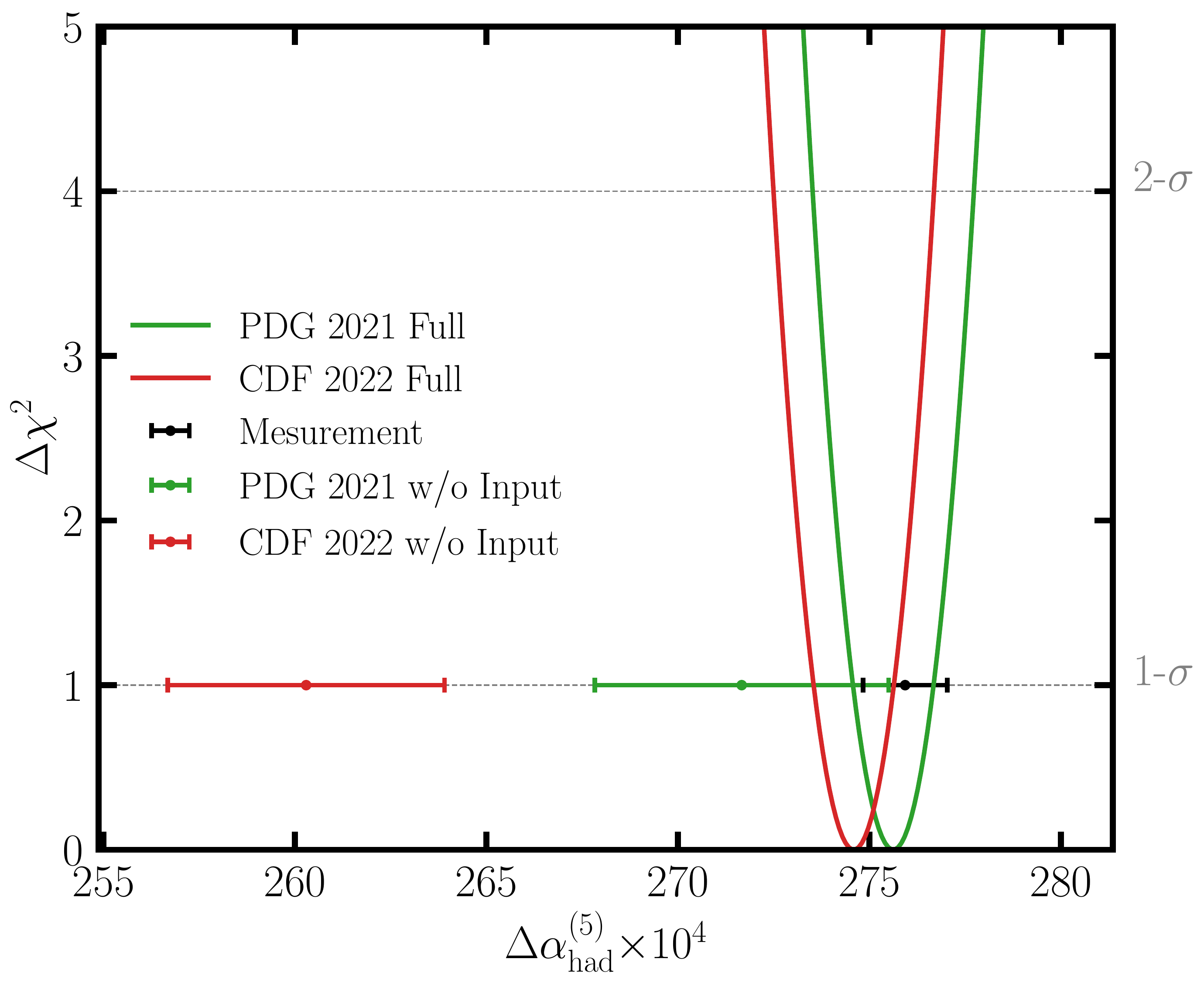}
\includegraphics[width=8cm,height=6cm]{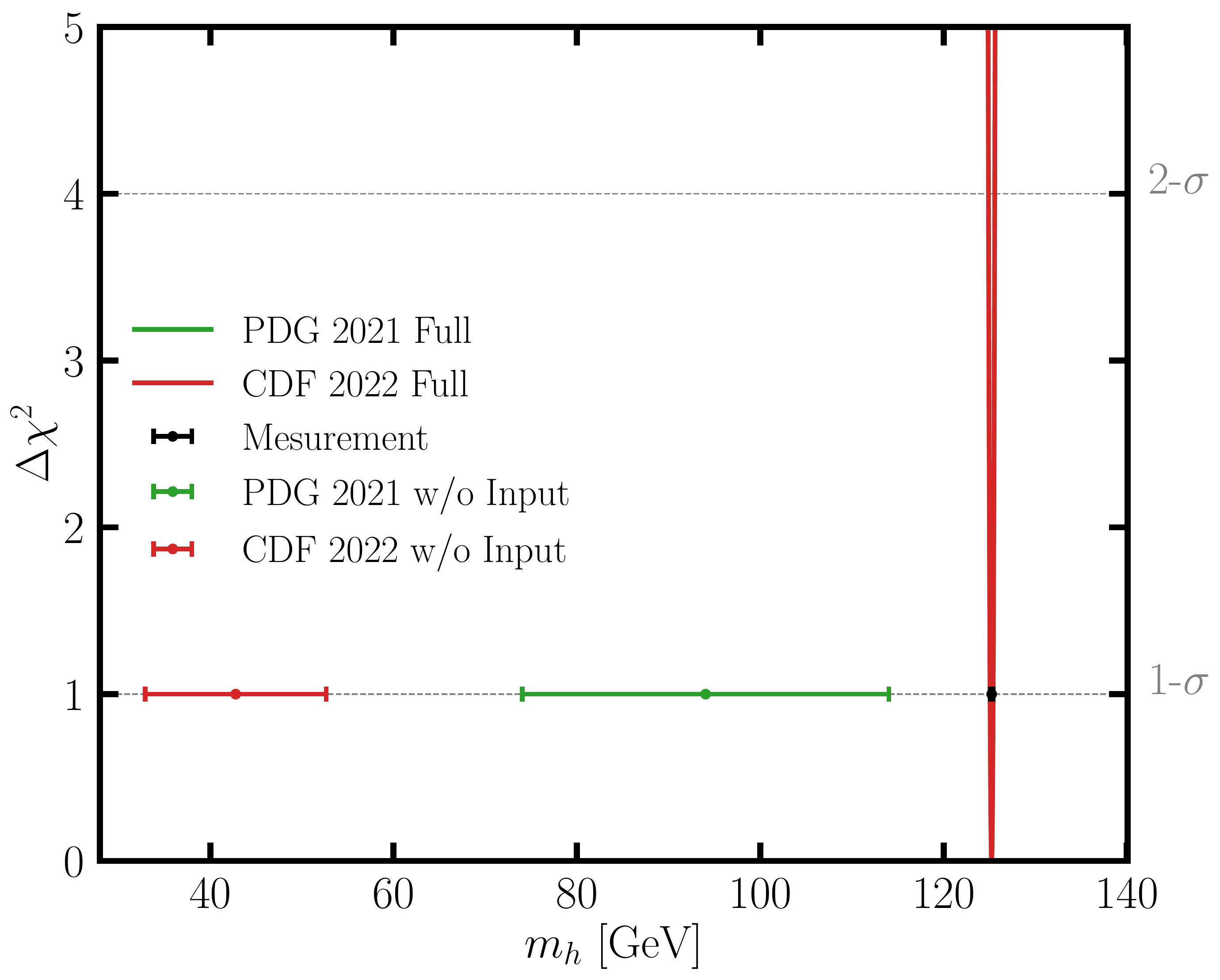}
\caption{Same as Fig.~\ref{fig:pull}, but for 1D EW fit results in $m_t$, $m_Z$, $\Delta\alpha^{(5)}_{\rm had} \times 10^4$ and $m_h$.}
\label{fig:1D_ewfit}
\end{figure}

\begin{figure*}[ht]
    \centering
    \includegraphics[width=16cm,height=6cm]{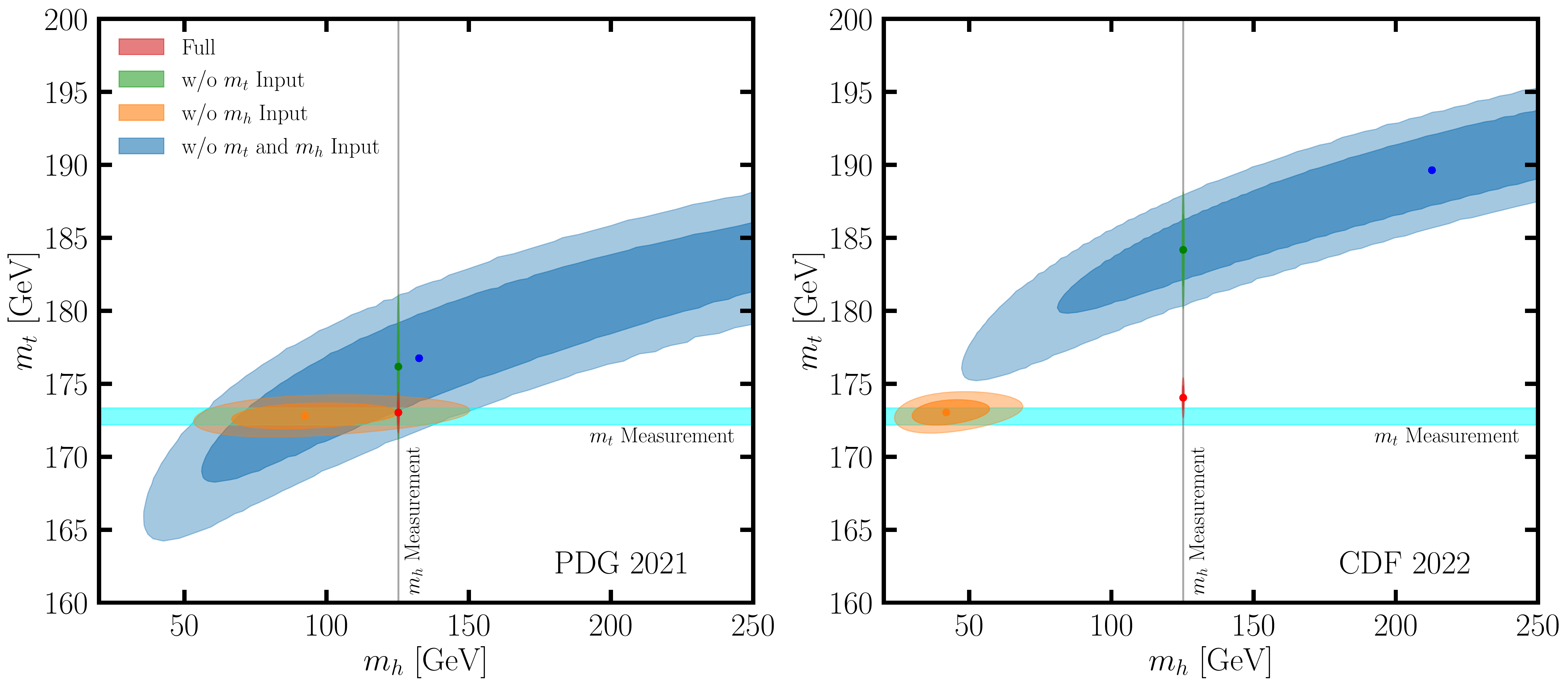}
    \caption{Same as Fig.~\ref{fig:pull}, but for 2D EW fit results in $m_t$-$m_h$ plane.}
\label{fig:2D_ewfit}
\end{figure*}

\begin{figure}[ht]
    \centering
    \includegraphics[width=0.4\textwidth,trim=10 50 10 0]{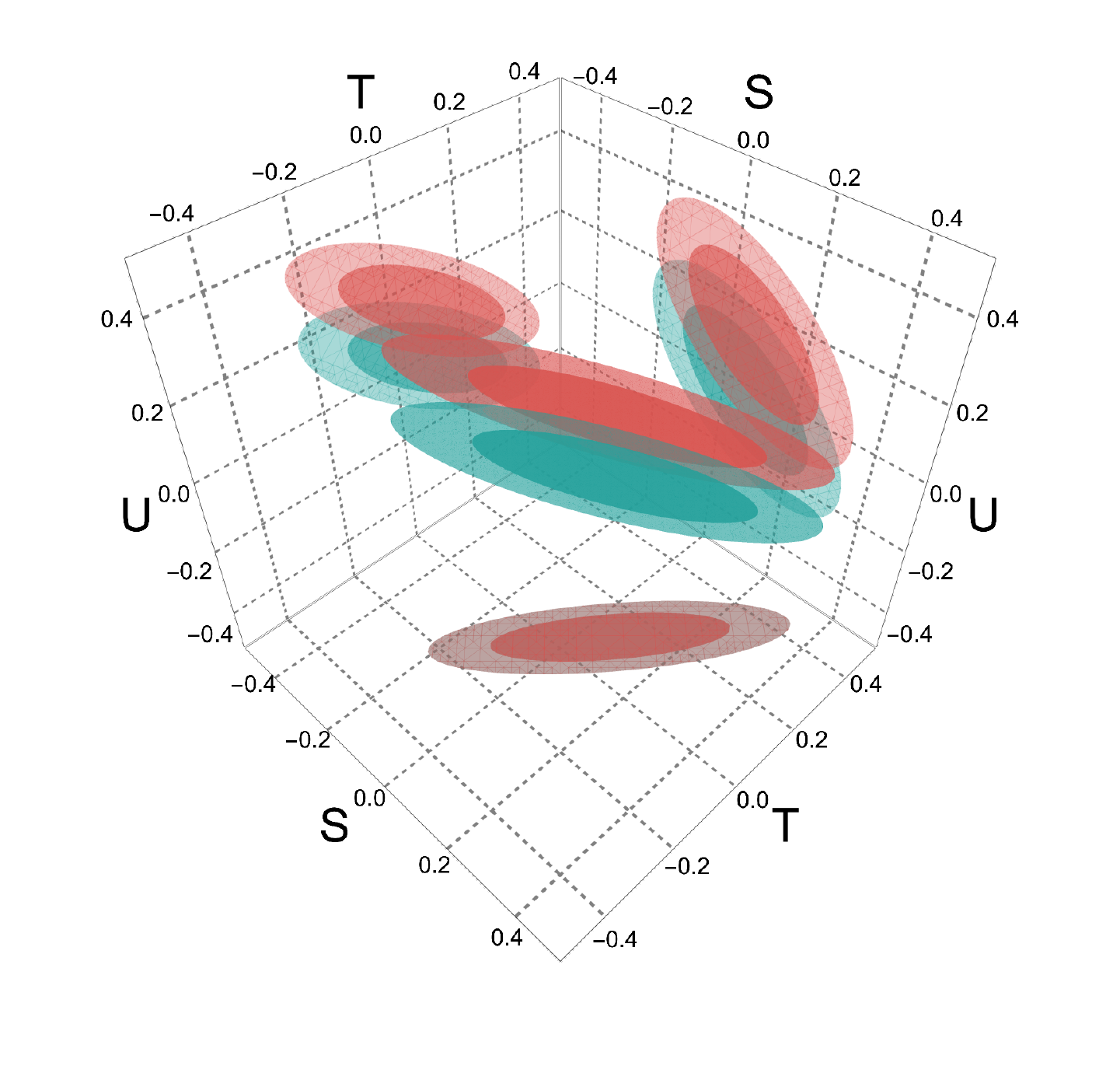}
    \caption{Same as Fig.~\ref{fig:pull}, but for 3D S/T/U fit results (old value of $m_W$: green region) and (new CDF value of $m_W$: red region) in $S$-$T$-$U$ ``space'' in the SM. The projections into individual two-dimension plane are also shown.}
    \label{fig:3D_ewfit}
\end{figure}

In this Supplemental Material, we visualize the numerical results from Tab.~\ref{tab:fit} in Fig.~\ref{fig:pull} and \ref{fig:1D_ewfit}, and also provide the two-dimensional fit results in $m_t$-$m_h$ plane and three-dimensional fit results in $S$-$T$-$U$ space for the SM in Fig.~\ref{fig:2D_ewfit} and~\ref{fig:3D_ewfit}, respectively. The analytic formulae and numerical results of the oblique parameters $S$, $T$ and $U$ in Two Higgs Doublet Model are given as well.

\ssection{Electroweak Fit Plots}
\label{ewplots}

\ssection{The oblique parameters in Two Higgs doublet Model}
\label{np}
The renormalizable and $CP$ invariant scalar potential for a general Two Higgs Doublet Model (THDM) with a softly broken $Z_2$ parity is given by,
\begin{equation}
\begin{aligned}
V\left(\Phi_{1}, \Phi_{2}\right)&=m_{11}^{2} \Phi_{1}^{\dagger} \Phi_{1}+m_{22}^{2} \Phi_{2}^{\dagger} \Phi_{2}-m_{12}^{2}\left(\Phi_{1}^{\dagger} \Phi_{2}+\Phi_{2}^{\dagger} \Phi_{1}\right)+\frac{\lambda_{1}}{2}\left(\Phi_{1}^{\dagger} \Phi_{1}\right)^{2}+\frac{\lambda_{2}}{2}\left(\Phi_{2}^{\dagger} \Phi_{2}\right)^{2} \\
&+\lambda_{3}\left(\Phi_{1}^{\dagger} \Phi_{1}\right)\left(\Phi_{2}^{\dagger} \Phi_{2}\right)+\lambda_{4}\left(\Phi_{1}^{\dagger} \Phi_{2}\right)\left(\Phi_{2}^{\dagger} \Phi_{1}\right)+\frac{\lambda_{5}}{2}\left[\left(\Phi_{1}^{\dagger} \Phi_{2}\right)^{2}+\left(\Phi_{2}^{\dagger} \Phi_{1}\right)^{2} \right]
\end{aligned}
\end{equation}
where $\Phi_{i=1,2}$ are two scalar $SU(2)$ doublets,
\begin{equation}
\Phi_{i}=\left(\begin{array}{c}
\phi_{i}^{+} \\
\left(v_{i}+\phi_{i}^{0}+i G_{i}^{0}\right) / \sqrt{2}
\end{array}\right).
\end{equation}
After the electroweak symmetry breaking, there are five physical Higgs bosons: two charged $H^{\pm}$ Higgs, two neutral $H$ and $h$ Higgs, and one neutral pseudoscalar $A$, whose masses are as follows:
\begin{equation}
\begin{aligned}
m_{H^{0}}^{2}=& \frac{m_{12}^{2}}{\sin \beta \cos \beta} \sin ^{2}(\beta-\alpha) \\
&+v^{2}\left[\lambda_{1} \cos ^{2} \alpha \cos ^{2} \beta+\lambda_{2} \sin ^{2} \alpha \sin ^{2} \beta+\frac{\lambda_{3}+\lambda_{4}+\lambda_{5}}{2} \sin 2 \alpha \sin 2 \beta\right] \\
m_{h^{0}}^{2}=& \frac{m_{12}^{2}}{\sin \beta \cos \beta} \cos ^{2}(\beta-\alpha) \\
&+v^{2}\left[\lambda_{1} \sin ^{2} \alpha \cos ^{2} \beta+\lambda_{2} \cos ^{2} \alpha \sin ^{2} \beta-\frac{\lambda_{3}+\lambda_{4}+\lambda_{5}}{2} \sin 2 \alpha \sin 2 \beta\right] \\
m_{A^{0}}^{2}=& \frac{m_{12}^{2}}{\sin \beta \cos \beta}-\lambda_{5} v^{2} \\
m_{H^{\pm}}^{2}=& \frac{m_{12}^{2}}{\sin \beta \cos \beta}-\frac{\lambda_{4}+\lambda_{5}}{2} v^{2}
\end{aligned}
\label{eqn:mass}
\end{equation}
The mixing of the neutral and charged Higgs fields is described by the angles $\alpha$ and $\beta$, which meet the following relations,
\begin{equation}
\begin{aligned}
\tan \beta &=\frac{v_{2}}{v_{1}}, \quad \tan 2 \alpha &=\frac{2\left(-m_{12}^{2}+\left(\lambda_{3}+\lambda_{4}+\lambda_{5}\right) v_{1} v_{2}\right)}{m_{12}^{2}\left(v_{2} / v_{1}-v_{1} / v_{2}\right)+\lambda_{1} v_{1}^{2}-\lambda_{2} v_{2}^{2}}
\end{aligned}
\end{equation}

The oblique parameters in 2HDM are given by,
\begin{equation}
\label{eq:stu-2hdm}
\begin{aligned}
\Delta S=& \frac{1}{\pi m_{Z}^{2}}\left\{\left[\mathcal{B}_{22}\left(m_{Z}^{2} ; m_{H}^{2}, m_{A}^{2}\right)-\mathcal{B}_{22}\left(m_{Z}^{2} ; m_{H^{\pm}}^{2}, m_{H^{\pm}}^{2}\right)\right]\right.\\
&+\left[\mathcal{B}_{22}\left(m_{Z}^{2} ; m_{h}^{2}, m_{A}^{2}\right)-\mathcal{B}_{22}\left(m_{Z}^{2} ; m_{H}^{2}, m_{A}^{2}\right)+\mathcal{B}_{22}\left(m_{Z}^{2} ; m_{Z}^{2}, m_{H}^{2}\right)-\mathcal{B}_{22}\left(m_{Z}^{2} ; m_{Z}^{2}, m_{h}^{2}\right)\right.\\
&\left.\left.-m_{Z}^{2} \mathcal{B}_{0}\left(m_{Z} ; m_{Z}, m_{H}^{2}\right)+m_{Z}^{2} \mathcal{B}_{0}\left(m_{Z} ; m_{Z}, m_{h}^{2}\right)\right] \cos ^{2}(\beta-\alpha)\right\}\\
\Delta T=& \frac{1}{16 \pi m_{W}^{2} s_{W}^{2}}\left\{\left[F\left(m_{H^{\pm}}^{2}, m_{A}^{2}\right)+F\left(m_{H^{\pm}}^{2}, m_{H}^{2}\right)-F\left(m_{A}^{2}, m_{H}^{2}\right)\right]\right.\\
&+\left[F\left(m_{H^{\pm}}^{2}, m_{h}^{2}\right)-F\left(m_{H^{\pm}}^{2}, m_{H}^{2}\right)-F\left(m_{A}^{2}, m_{h}^{2}\right)+F\left(m_{A}^{2}, m_{H}^{2}\right)\right.\\
&+F\left(m_{W}^{2}, m_{H}^{2}\right)-F\left(m_{W}^{2}, m_{h}^{2}\right)-F\left(m_{Z}^{2}, m_{H}^{2}\right)+F\left(m_{Z}^{2}, m_{h}^{2}\right) \\
&\left.\left.+4 m_{Z}^{2} \bar{B}_{0}\left(m_{Z}^{2}, m_{H}^{2}, m_{h}^{2}\right)-4 m_{W}^{2} \bar{B}_{0}\left(m_{W}^{2}, m_{H}^{2}, m_{h}^{2}\right)\right] \cos ^{2}(\beta-\alpha)\right\}\\
\Delta U=&-\Delta S+\frac{1}{\pi m_{W}^{2}}\left\{\left[\mathcal{B}_{22}\left(m_{W}^{2}, m_{A}^{2}, m_{H^{\pm}}^{2}\right)-2 \mathcal{B}_{22}\left(m_{W}^{2}, m_{H^{\pm}}^{2}, m_{H^{\pm}}^{2}\right)+\mathcal{B}_{22}\left(m_{W}^{2}, m_{H}^{2}, m_{H^{\pm}}^{2}\right)\right]\right.\\
&+\left[\mathcal{B}_{22}\left(m_{W}^{2}, m_{h}^{2}, m_{H^{\pm}}^{2}\right)-\mathcal{B}_{22}\left(m_{W}^{2}, m_{H}^{2}, m_{H^{\pm}}^{2}\right)+\mathcal{B}_{22}\left(m_{W}^{2}, m_{W}^{2}, m_{H}^{2}\right)-\mathcal{B}_{22}\left(m_{W}^{2}, m_{W}^{2}, m_{h}^{2}\right)\right.\\
&\left.\left.-m_{W}^{2} \mathcal{B}_{0}\left(m_{W}^{2}, m_{W}^{2}, m_{H}^{2}\right)+m_{W}^{2} \mathcal{B}_{0}\left(m_{W}^{2}, m_{W}^{2}, m_{h}^{2}\right)\right] \cos ^{2}(\beta-\alpha)\right\}
\end{aligned}
\end{equation}
where the loop functions are provided in Ref.~\cite{tHooft:1978jhc} and can be numerically calculated with {\tt Looptools}~\cite{Hahn:1998yk}.

\begin{figure}[ht]
\includegraphics[width=5.4cm,height=4.8cm]{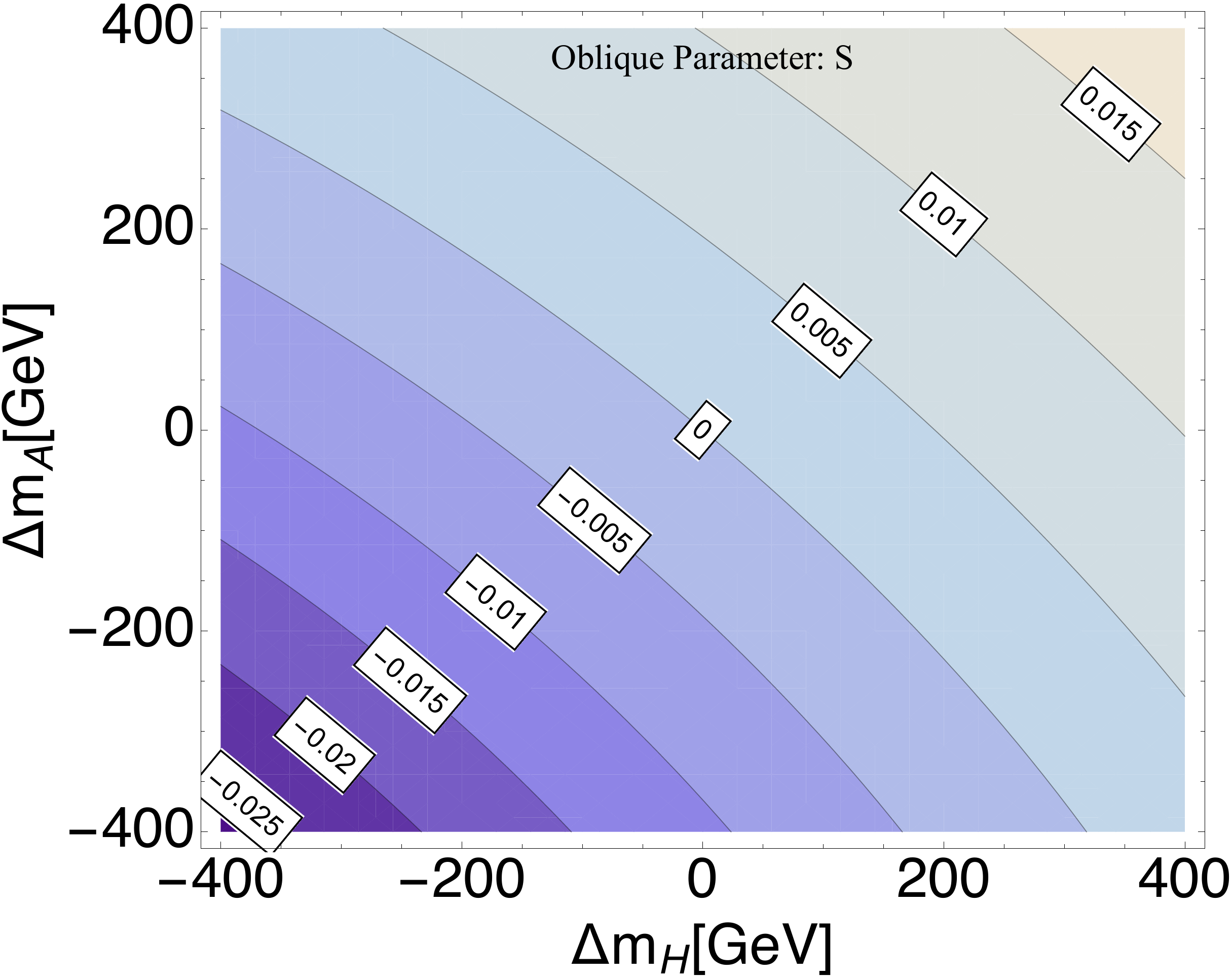}
\includegraphics[width=5.4cm,height=4.8cm]{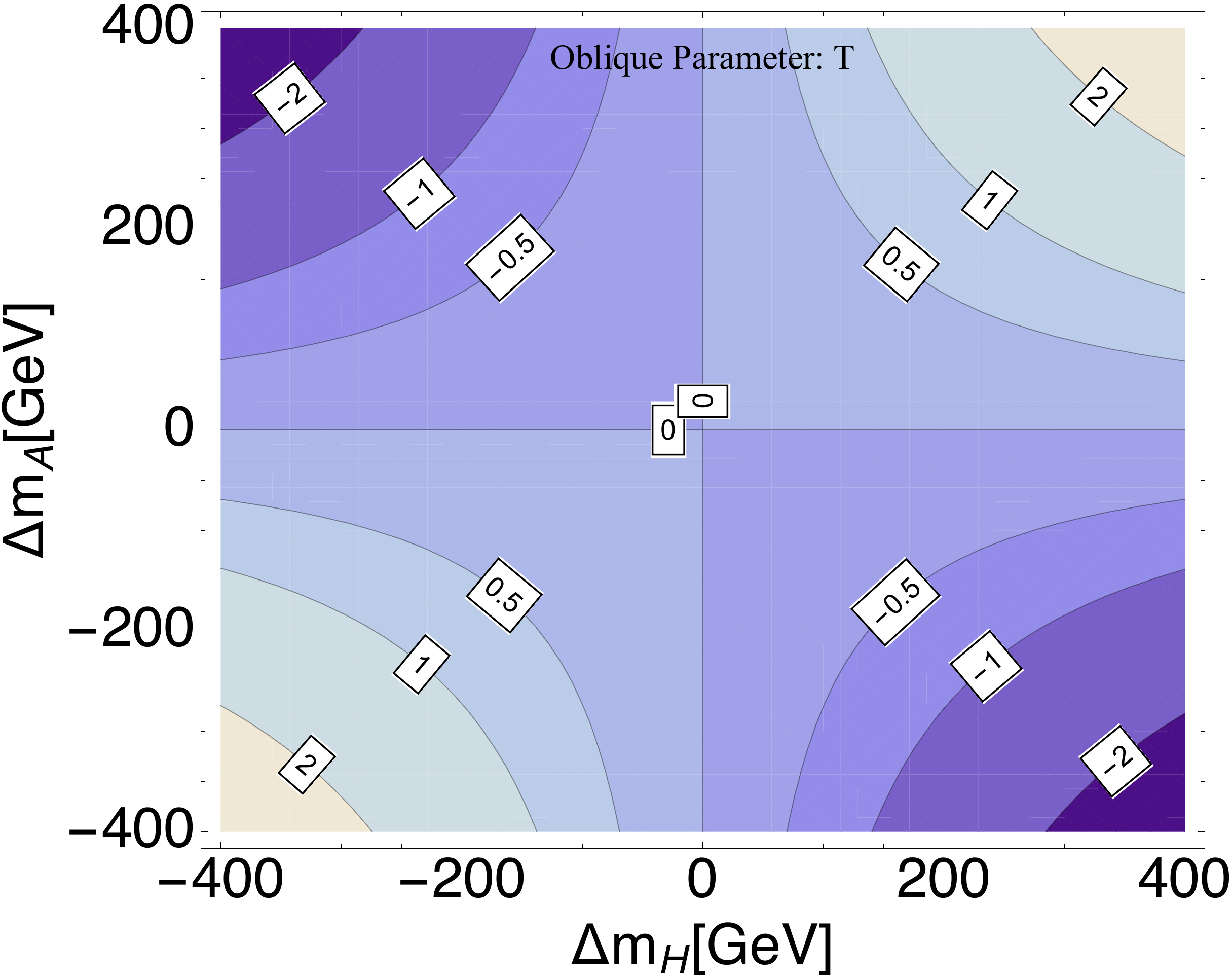}
\includegraphics[width=5.4cm,height=4.8cm]{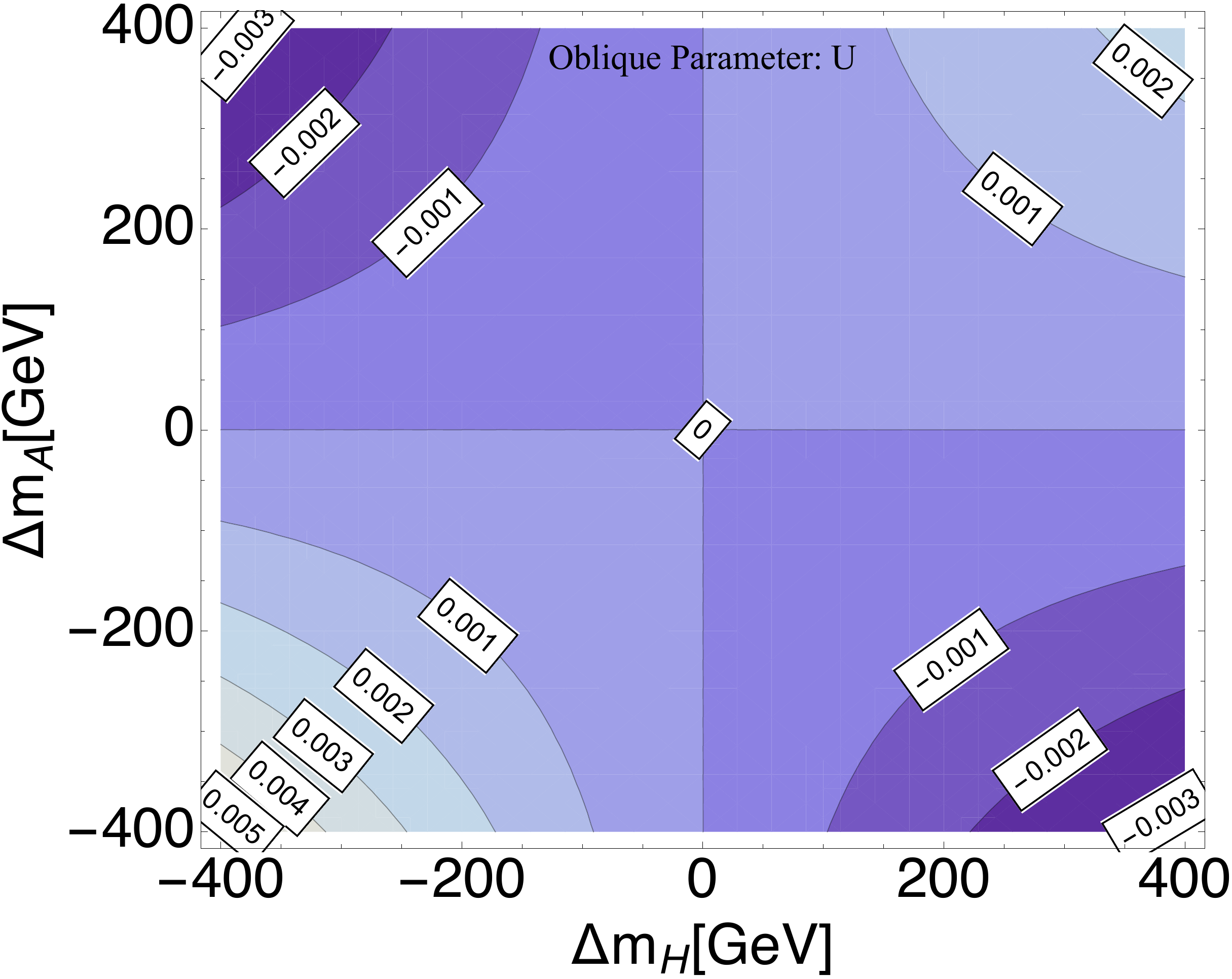}
\caption{Contour plots of $S$, $T$ and $U$ for $m_{H^{\pm}}=1$ TeV and $\cos(\beta-\alpha)=0$ in $\Delta m_H$ and $\Delta m_A$ plane.}
\label{fig:2hdm_stu_0}
\end{figure}

\begin{figure}[ht]
\includegraphics[width=5.4cm,height=4.8cm]{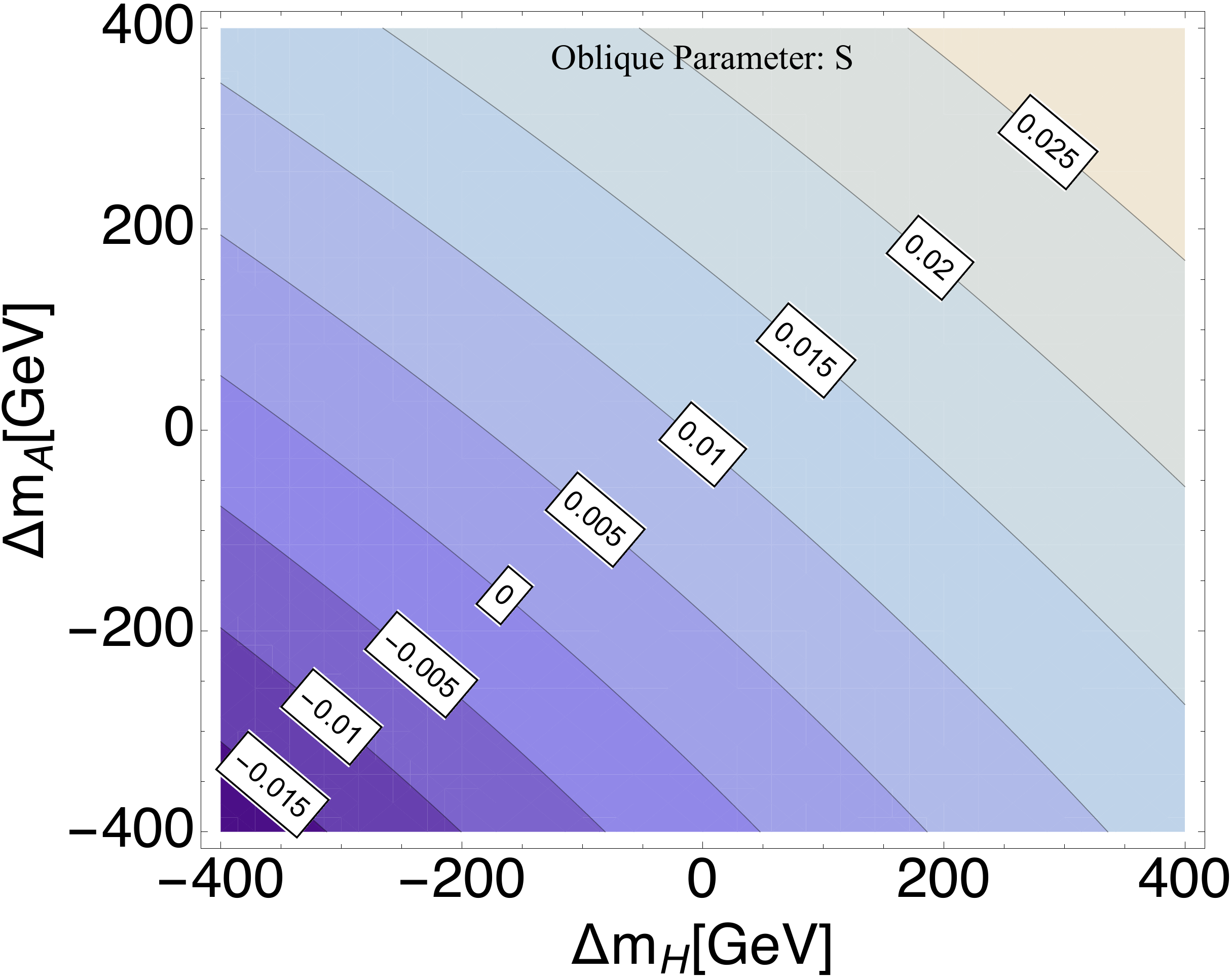}
\includegraphics[width=5.4cm,height=4.8cm]{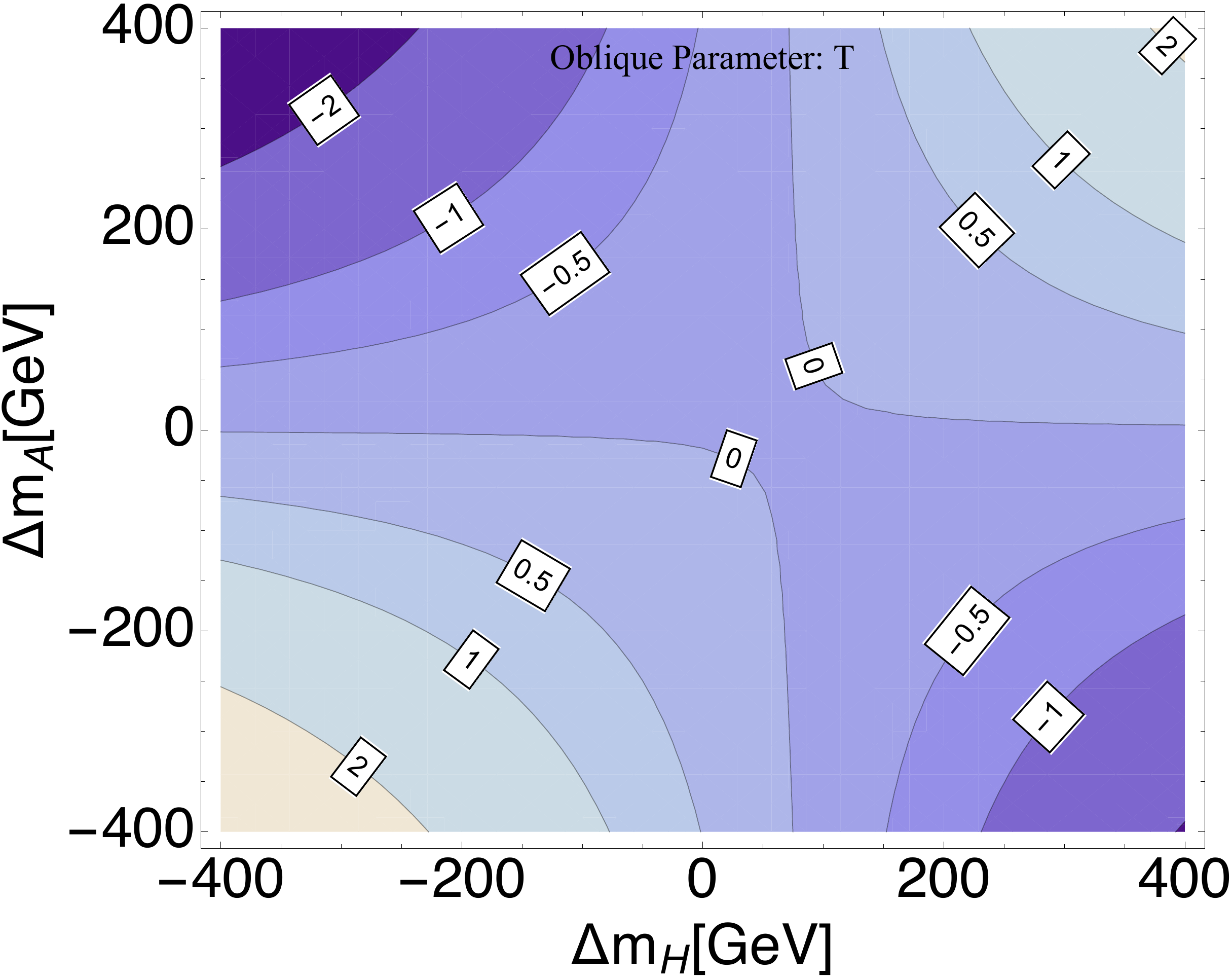}
\includegraphics[width=5.4cm,height=4.8cm]{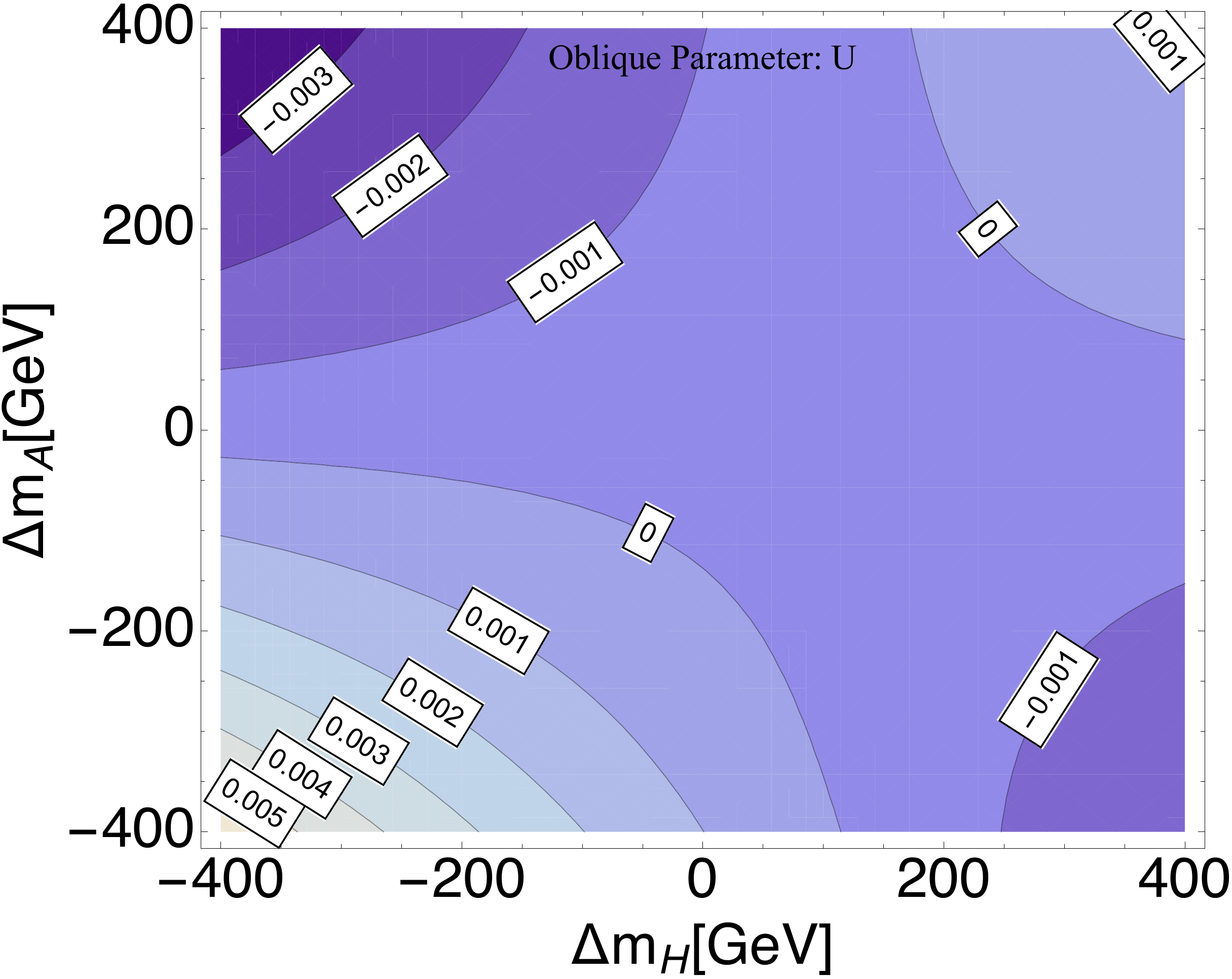}
\caption{Same as Fig.~\ref{fig:2hdm_stu_0}, but for $\cos(\beta-\alpha)=0.3$.}
\label{fig:2hdm-stu-1}
\end{figure}

\end{document}